\documentclass[prd,floatfix,nofootinbib,preprintnumbers]{revtex4}
\usepackage{epsfig}
\usepackage{graphicx}
\usepackage{amsfonts}
\usepackage{mathrsfs}

\begin{document}

\newcommand{\be}{\begin{equation}}
\newcommand{\ee}{\end{equation}}
\newcommand{\bea}{\begin{eqnarray}}
\newcommand{\eea}{\end{eqnarray}}
\newcommand{\nnb}{\nonumber}
\renewcommand{\thefootnote}{\fnsymbol{footnote}}
\def\lsim{\raise0.3ex\hbox{$\;<$\kern-0.75em\raise-1.1ex\hbox{$\sim\;$}}}
\def\gsim{\raise0.3ex\hbox{$\;>$\kern-0.75em\raise-1.1ex\hbox{$\sim\;$}}}
\def\Frac#1#2{\frac{\displaystyle{#1}}{\displaystyle{#2}}}
\def\no{\nonumber\\}
\def\slash#1{\ooalign{\hfil/\hfil\crcr$#1$}}
\def\ep{\eta^{\prime}}
\def\susy{\mbox{\tiny SUSY}}
\def\sm{\mbox{\tiny SM}}
\def\pslash{\rlap{\hspace{0.02cm}/}{p}}
\def\qslash{\rlap{/}{q}}
\def\kslash{\rlap{\hspace{0.02cm}/}{k}}
\def\lslash{\rlap{\hspace{0.011cm}/}{\ell}}
\def\nslash{\rlap{\hspace{0.02cm}/}{n}}
\def\Pslash{\rlap{\hspace{0.065cm}/}{P}}
\textheight      250mm  

\vskip0.5pc

\title{CP asymmetry in $B\to \phi K_S$ in a SUSY SO(10) GUT}

\author{
Yuan-Ben Dai$^{a}$,  Chao-Shang Huang$^a$, Wen-Jun Li$^a$, and Xiao-Hong Wu$^b$
}

\affiliation{
 $^a$ Institute of Theoretical Physics, Academia Sinica, P. O. Box 2735,
             Beijing 100080,  China\\
 $^b$ Department of Physics, KAIST, Daejeon 305-701, Korea
}

\begin{abstract}
We study the $B\to \phi K_S$  decay in a SUSY SO(10) GUT. We
calculate the mass spectrum of sparticles for a given set of
parameters at the GUT scale. We complete the calculations of the
Wilson coefficients of operators including the new operators which
are induced by NHB penguins at LO using the MIA with double
insertions. It is shown that the recent experimental results on
the time-dependent CP asymmetry $S_{\phi K}$ in $B\to \phi K_S$,
which is negative and can not be explained in SM, can be explained
in the model where there are flavor non-diagonal right-handed down
squark mass matrix elements of 2nd and 3rd generations whose size
satisfies all relevant constraints from known experiments ($\tau
\to \mu \gamma$, $B\to X_S\gamma, B_s\to \mu^+\mu^-, B\to X_s
\mu^+\mu^-, B\to X_s g, \Delta M_s$, etc.). At the same time, the
branching ratio for the decay can also be in agreement with
experimental measurements.
\end{abstract}

\maketitle
\noindent

\section{Introduction}
Great progresses have been made on the flavor physics in recent
years. Among them the progress in neutrino physics is particularly
impressive. Atmospheric neutrino~\cite{skatm} and solar
neutrino~\cite{sno} experiments together with the reactor
neutrino~\cite{kamland,chooz} experiments have established the
oscillation solution to the solar and atmospheric neutrino
anomalies, which signals the existence of new physics beyond the
SM. Experiment results indicate smallness of the masses of
neutrinos and the bilarge mixing pattern among the three
generations of neutrinos. Because the small quark mixing in the
CKM matrix is related to the large quark mass
hierarchy~\cite{wwzf}, understanding the bilarge mixing pattern is
somewhat of a challenge. However, if allowing asymmetric form for
the mass matrix which for example may well be generated by the
elegant Froggatt-Nielsen (FN) mechanism~\cite{FN}, we can
accommodate the large mass hierarchy with large mixing~\cite{asy}.
Therefore, if one works with an effective theory, {\it e.g.}, the
minimal supersymmetric Standard Model together with right-hand
neutrinos (MSSM+N), at a low energy scale (say, the electro-weak
scale), one can content oneself by using the WWZF scenario to
understand the smallness of quark mixing, and the see-saw and FN
mechanism to understand the smallness of the masses of neutrinos
and the largeness of neutrino mixing. However, if one works with a
theory, {\it e.g.}, a grand unification theory (GUT), in which
quarks and leptons are in a GUT multiplet, one has to answer: can
we explain simultaneously the smallness of quark mixing and the
largeness of neutrino mixing in the theory? If we can, then what
are the phenomenological consequences in the theory? There are
several recent works to tackle these problems in SU(5), flipped
SU(5), or SO(10) GUTs ~\cite{asy,bdv,lfv3,cmm,mvv,bsv,hs,hll},
which brings the study of GUT to a more realistic level.

The smallest grand unification group that incorporates righthanded
neutrinos required for the seesaw mechanism, is SO(10). It has
been shown that the observed bilarge mixing naturally leads to
large flavor non-diagonal down-type squark mass matrix elements of
2nd and 3rd generations in a SUSY SO(10)~\cite{cmm}.

The measurements of the time dependent CP asymmetry $S_{J/\psi K}$
in $B\to J/\psi K_S$ have established the presentence of CP
violation in neutral B meson decays and the measured
value\cite{sj} \be S_{J/\psi K}=\sin (2 \beta (J/\psi K_{S}))_{\rm
world-ave}=0.734 \pm 0.054. \label{sjp}\ee is in agreement with
the prediction in the standard model (SM). Recently, various
measurements of  CP violation in B factory experiments have
attracted much interest. Among them\cite{2002,2003},
\begin{eqnarray} S_{\phi K_S}&=&-0.39\pm 0.41,\,\,\,\,\,2002\,\,\,
{\rm World-average}\nnb\\S_{\phi K_S}&=&-0.15\pm 0.33,\,\,\,\,\,
2003\,\,\, {\rm World-average}
\end{eqnarray} is
especially interesting since it deviates greatly from the SM
expectation \be S_{\phi K_S}=\sin (2 \beta (\phi K_S))= \sin (2
\beta (J/\psi K_S)\!) \! + \! O(\lambda^2 \!)
\end{equation}
where $\lambda \simeq 0.2$ appears in Wolfenstein's
parameterization of the CKM matrix. Though the impact of these
experimental results on the validity of CKM and SM is currently
limited by experimental uncertainties, they have attracted much
interest in searching for new physics~\cite{dat,kk,kane,chw,hz}
and it has been shown that the deviation can be understood without
contradicting the smallness of the SUSY effect on $B \to J/\psi
K_S$ in the minimal supersymmetric standard model
(MSSM)~\cite{kane,chw}. Motivated by SUSY GUTs, the time dependent
CP asymmetry in $B\to \phi K_S$ has been studied in the SUSY
models with a large mixing between the 2nd and 3rd generation in
the down-type squark mass matrix at the $M_{SUSY}\sim TeV$
scale~\cite{hlm}. The asymmetry has also been examined in the SUSY
SU(5) framework~\cite{cmsvv}. It is shown in ref. \cite{cmsvv}
that the possibility of large deviations from SM in $S_{\phi K}$
is excluded in the case of only having the flavor non-diagonal
down-type squark mass matrix element of 2nd and 3rd generations in
the RR sector (i.e., $\delta_{23}^{dRR}$ non-zero with
$\delta_{23}^{dRR}\equiv (M^2_{\tilde{d}RR})_{23} /
m_{\tilde{q}}^2$, where $(M^2_{\tilde{d}RR})_{ij}$ is the flavor
non-diagonal squared right-handed down squark mass matrix element
and $m_{\tilde{q}}$ is the average right-handed down-type squark
mass) due to the bound on $\delta_{23}^{dRR}$ from $Br(\tau \to
\mu \gamma)$. In this paper we investigate the decay $B\to \phi
K_S$ in the SUSY SO(10) framework. Our results show that the time
dependent CP asymmetry in $B\to \phi K_S$ can sizably deviate from
SM after imposing the constraint from $Br(\tau \to \mu \gamma)$,
in contrast with the claim in the literature, because we have
included the double insertion contributions in penguin diagrams
for the relevant Wilson coefficients. Furthermore, the
contributions from neutral Higgs penguins, which are not
considered in ref. \cite{cmsvv}, have been included in this paper.

We need to have new CP violation sources in addition to that of
CKM matrix in order to explain the deviations of $S_{\phi K_S}$
from SM. There are new sources of flavor and CP violation in SUSY
SO(10) GUTs. In such kind of models there is a complex flavor
non-diagonal down-type squark mass matrix element of 2nd and 3rd
generations of order one at the GUT scale~\cite{cmm} which can
induce large flavor off-diagonal couplings such as the coupling of
gluino to the quark and squark which belong to different
generations. These couplings are in general complex and
consequently can induce CP violation in flavor changing neutral
currents (FCNC). It is well-known that the effects of the
counterparts of usual chromo-magnetic and electro-magnetic dipole
moment operators with opposite chirality are suppressed by
${m_s}/{m_b}$ and consequently negligible in SM. However, in SUSY
SO(10) GUTs their effects can be significant, since
$\delta_{23}^{dRR}$ can be as large as 0.5~\cite{cmm}.

For the $b\to s$ transition, besides the SM contribution, there
are mainly two new contributions arising from the QCD and
chromo-magnetic penguins and neutral Higgs boson (NHB) penguins
with the sparticles propagating in the loop in SUSY models. The
contributions to the relevant Wilson coefficients at the $m_W$
scale from the diagram with the gluino propagating in the loop
have been calculated by using the mass insertion approximation
(MIA)~\cite{mi} with double insertions in ref.~\cite{chw}. We
calculate the contributions from the diagram with the chargino
propagating in the loop using MIA with double insertions because
chargino contributions can be significant when the left-right
mixing between squark masses is large.

As it is shown that both Br and CP asymmetries depend
significantly on how to calculate hadronic matrix elements of
local operators\cite{chw}. Recently, two groups, Li et al.
\cite{li1,li} and BBNS~\cite{bbns,bbns1}, have made  significant
progress in calculating hadronic matrix elements of local
operators relevant to charmless two-body nonleptonic decays of B
mesons in the PQCD framework. The key point to apply PQCD is to
prove that the factorization, the separation of the short-distance
dynamics and long-distance dynamics, can be performed for those
hadronic matrix elements. It has been shown that in the heavy
quark limit (i.e., $m_b\to\infty$) such a separation is indeed
valid and hadronic matrix elements can be expanded in $\alpha_s$
such that the tree level (i.e., the $\alpha_s^0$ order) is the
same as that in the naive factorization and the $\alpha_s$
corrections can be systematically calculated\cite{bbns}. Comparing
with the naive factorization, to include the $\alpha_s$ correction
decreases significantly the hadronic uncertainties. In particular,
the matrix elements of the chromomagnetic-dipole operators
$Q_{8g}^{(\prime)}$ have large uncertainties in the naive
factorization calculation which lead to the significant
uncertainty of the time dependent CP asymmetry in SUSY
models\cite{hlm}. The uncertainties are greatly decreased in BBNS
approach\cite{bbns1}. The hadronic matrix element of operators
relevant to the decays $B\rightarrow \phi K_S$ up to the
$\alpha_s$ order have been calculated in BBNS approach in
ref.~\cite{chw}.

Using the BBNS approach to calculate hadronic elements to the
$\alpha_s$ order and Wilson coefficients in MIA with double
insertions, we show in this paper that in the SUSY $SO(10)$
framework in the reasonable region of parameters where the
constraints from $\tau\to\mu\gamma$, $B_s-\bar{B}_s$ mixing ,
$\Gamma(b \to s \gamma)$, $\Gamma(b \to s g)$, $\Gamma(b \to
s\mu^+ \mu^-)$, and $B\to \mu^+\mu^-$ are satisfied, the branching
ratio of the decay for $B\rightarrow \phi K_S$ can be smaller than
$1.6 \times 10^{-5}$, and the theoretical prediction for $S_{\phi
K}$ can be in agreement with the data in $1 \sigma$ experimental
bounds and even can be as low as $-0.6$.

The paper is organized as follows. In Section II we describe the
SO(10) models we used in the paper briefly. In section III we
calculate sparticle spectrum using revised ISAJET. In section IV
we give the effective Hamiltonian responsible for $B\rightarrow
\phi K_S$ in the model. In particular, we give the Wilson
coefficients of operators using MIA with double insertions.
 The Section V is devoted to numerical results of the
time dependent CP asymmetry and branching ratio for the decay
$B\to\phi K_S$. We draw conclusions and discussions in Section VI.

\section{Neutrino Bilarge Mixing and large b-s transitions in SO(10)GUT}
It is well-known for a long time that SO(10) GUTs can naturally
incorporate the seesaw mechanism~\cite{seesaw} which is the
simplest way to understand small neutrino masses. Recently a
number of SUSY SO(10) models which can explain neutrino data and
quark mixing have been proposed~\cite{asy,cmm,bsv} and some
phenomenological consequences of the models have been
analyzed~\cite{asy,bi,cmm}. It was first pointed out in
~\cite{cmm} that the induced large mixing in the down-type squark
mass matrix in SUSY $SO(10)$ models have interesting consequences
in low energy B physics. For specific, we use the model in
ref.~\cite{cmm} and review the main points of the model. The
details of the model can be found in Ref.~\cite{cmm}.

In order to accommodate both CKM mixing among quarks and MNS
mixing among leptons, one needs to have an asymmetric down-type
Yukawa matrix $Y_d$. So instead of the usual superpotential
\cite{dh}, one assumes
\begin{equation}
  W = \frac{1}{2} (Y_u)_{ij} 16_i 16_j 10_u
  + \frac{1}{2} (Y_d)_{ij} 16_i 16_j \frac{\langle 45 \rangle}{M_{Pl}}
  10_d.
\end{equation}
Because of the combination of the Higgs multiplet $45$, whose VEV
$\langle 45 \rangle \neq 0$ breaks SO(10), and the Higgs in 10,
the effective Yukawa coupling being either in 10 (symmetric
between two 16's) or 120 (anti-symmetric between two 16's)
representations, the matrix $Y_d$ can now have a mixed symmetry.
Setting the breaking chain to be $SO(10)\to SU(5)\to SM$, we have
the Yukawa couplings in the MSSM+N (the MSSM with right-handed
neutrinos) as
\begin{eqnarray}
  \lefteqn{
    W = (Y_u^D)_{i} Q_i U_i H_u  + (Y_u^D)_{i} L_i N_i H_u
  }
  \nonumber \\
  & & \hspace{-0.5cm}
+ (V^* Y_d^D U )_{ij} Q_i D_j H_d + (V^* Y_d^D U )_{ij} E_i L_j
H_d
  + \frac{1}{2} M_{ij} N_i N_j,
  \label{eq:MSSM+N}
\end{eqnarray}
where $Y_i^D$ (i=u, d) is the diagonal and we have absorbed the
phase matrix $\Theta_L$ into $Q_i$ and $E_i$,  the phase matrix
$\Theta_R$ into $(UD)$ multiplet or $(EV^*)$ multiplet, and the
phase matrix $\Theta_\nu$ into the Majorana mass matrix $M$,
respectively. Here V and U are the usual CKM mixing matrix and
neutrino mixing matrix (MNS matrix) respectively. We have taken
{\it $Y_u$ and $M$ simultaneously diagonal}\/. Such a situation
could result from simple $U(1)$ family symmetries and in a
$SO(10)$ with hierarchical $Y_u^D$ and right-handed neutrino
masses, the choice of having such simultaneous diagonalization
looks rather plausible~\cite{bw}.

After integrating out the right-handed neutrinos in
Eq.~(\ref{eq:MSSM+N}), the light neutrino masses are determined
from the superpotential
\begin{equation}
  W = \frac{1}{2} (Y_u^D)_{i} (M^{-1})_{ij} (Y_u^D)_{j} (L_i H_u) (L_j
  H_u),
\end{equation}
which leads to the light Majorana neutrino mass matrix
$(m_\nu)_{nm} = (Y_{ui}^{D 2}/2M_i) U_{ni}^* e^{- i \delta_i}
U_{mi}^*$, where $e^{i \delta_i}$ is the phase of the diagonal
element $M_i$ of the diagonal matrix M. Therefore, the two mass
splitting data and bilarge mixing of neutrinos can be explained.

The Yukawa coupling of ``third-generation'' neutrino is unified
with the large top Yukawa coupling due to the $SO(10)$
unification. Nevertheless, this ``third-generation'' neutrino is
actually a near-maximal mixture of $\nu_\mu$ and $\nu_\tau$, which
comes from the large mixing angle in atmospheric neutrino
oscillation. Because the ``third-generation'' charged leptons,
down-type quarks and neutrinos are in a $SU(5)$ multiplet, the
$SU(5)$ multiplet with the large top Yukawa coupling contains
approximately
\begin{equation}
  5^*_3 = 5^*_\tau \cos \theta + 5^*_\mu \sin \theta,
\end{equation}
where $\theta \simeq 45^\circ$ is the atmospheric neutrino mixing
angle, and
\begin{eqnarray}
  5^*_\tau &=& (b^c, b^c, b^c, \nu_\tau, \tau)\\
  5^*_\mu  &=& (s^c, s^c, s^c, \nu_\mu,  \mu).
\end{eqnarray}
The similar mixing exists in right-handed b and s quarks. However,
we do not worry about its effects because there is no
charged-current weak interaction on right-handed quarks. Therefore
the mixing among right-handed quarks decouples from low-energy
physics.

We now come to the main outcome in the framework which has
phenomenological consequences in low-energy B physics. Because one
works in SUSY models there is the corresponding mixing among
squarks which would yield observable effects. The top Yukawa
coupling generates an $O(1)$ radiative correction to the mass of
$\tilde{s}\sin\theta+\tilde{b}\cos\theta$, which leads to a large
mixing between $\tilde{s}$ and $\tilde{b}$ at low energies. This
large mixing in turn generates interesting effects in $B$-physics.

To be specific, assuming that at the scale $M_*$ which is above
the GUT unification scale, say, near the Planck scale, one has the
universal soft terms: scalar mass $m_0$, gaugino mass $M_{1/2}$,
trilinear and bilinear couplings $A_0$ and $B_0$, then large
neutrino Yukawa couplings involved in the neutrino Dirac masses
can induce large off-diagonal mixing in the right-handed down
squark mass matrix through renormalization group evolution between
$M_*$ and $M_{GUT}$ and the induced mixings will generally be
complex with new CP violating phases. The induced off-diagonal
elements in the mass matrix of the right-handed down squarks
$\tilde{d}_R$ are given by (in the basis in which $Y_D$ is
diagonal) \be [m^2_{\tilde{d}_R}]_{nm} \simeq -\frac{1}{8\pi^2}
\left[ Y^{u\dagger} Y^u \right]_{nm} (3m_0^2+a_0^2) \left(5 \log
\frac{M_*}{M_{10}} + \log \frac{M_{10}}{M_{5}} \right)
\label{eq:msquare} \ee where $M_5$ is the $SU(5)$ breaking scale
and
\begin{equation}\label{yu23}
\left[ Y^{u\dagger} Y^u \right]_{nm} = \left[ (\Theta_R U Y_u^{D
2} U^\dagger \Theta_R^* \right]_{nm} =
e^{-i(\phi^{(L)}_m-\phi^{(L)}_n)} y_t^2 \left[U\right]_{m3}^*
\left[U\right]_{n3},
\end{equation}
where $e^{i \phi^{(L)}_n}$ is the phase from $(\Theta_R)_{nn}$.
Note that these phases are not relevant to any other low energy
physics.

Substituting $U_{23,33}\sim \sqrt{2}$ into eq.(\ref{yu23}), one
obtains
\begin{equation}
[Y^{u\dagger}Y^u]_{23} = 0.5
e^{-i(\phi^{(L)}_2-\phi^{(L)}_3)}(m_{t}(M_G)/178\mbox{ GeV}) ^2,
\label{eq:delta} \ee where $m_{t}(M_G)$ is the top quark mass at
$M_{G}$. From eqs. (\ref{eq:msquare},\ref{eq:delta}), the mass
insertion parameter $\delta^{dRR}_{23}$ can easily be of order
one.

\section{Mass spectra and the permitted parameter space}
To see the impact of the induced off-diagonal elements in the mass
matrix of the right-handed down-type squarks on B rare decays and
simplify the analysis, we assume that at the GUT scale ($M_G$) all
sfermion mass matrices except the right-handed down-type squark
mass matrix are flavor diagonal and all diagonal elements are
approximately universal and equal to $m^2_0$. The 2-3 matrix
element of right-handed down-type squark mass matrix is
parameterized by $\delta_{23}^{dRR}\equiv
(M^2_{\tilde{d}RR})_{23}/m_0^2$ which can be treated as a free
parameter of order one, as discussed in last section. Furthermore,
we have a universal gaugino mass $M_{1/2}$, a universal trilinear
coupling $A_0$ and a universal bilinear coupling $B_0$ at $M_G$.
With the requirement of radiative electro-weak (EW) symmetry
breaking, we have five parameters ($m_0, M_{1/2}, A_0,
\delta_{23}^{dRR}, tan\beta$) plus the sign of $\mu$ as the
initial conditions for solving the renormalization group equations
(RGEs).

We require that the lightest neutralino be the lightest
supersymmetric particle (LSP) and use several experimental limits
to constraint the parameter space, including 1)the width of the
decay $Z \rightarrow \chi^0_1 \chi^0_1$ is less than 4.3 MeV, and
branching ratios of $Z \rightarrow \chi^0_1 \chi^0_2$ and $Z
\rightarrow \chi^0_2 \chi^0_2$ are less than $1 \times 10^{-5}$,
where $\chi^0_1$ is the lightest neutralino and $\chi^0_2$ is the
other neutralino, 2) the mass of light neutral even Higgs can not
be lower than 111 GeV as the present experments required, 3) the
mass of lighter chargino must be larger than 94 GeV as given by
the Particle Data Group~\cite{pdg}, 4) sneutrinos are larger than
94GeV, 5) seletrons are larger than 73GeV, 6) smuons larger than
94 GeV, 7) staus larger than 81.9 GeV.

We use revised ISAJET to do numerical calculations. We find that
the parameter
$(M^2_{\tilde{d}RR})_{23}$
does not receive any significant
correction and the diagonal entries are significantly corrected,
which is in agreement with the results in ref.~\cite{cmsvv}. We
scan $m_0,\;M_{1/2}$ in the range (100, 800) GeV for given values
of $A_0, \tan\beta$ and sign($\mu$)=+1\footnote{In the case of
sign($\mu$)= -1, the constraint from $B\to X_s \gamma$ on the
parameter space is too stringent, in particular, for large
$\tan\beta$~\cite{nanop,hy}.} We impose the constraints from the
relevant low energy experiments such as $B\to X_s \gamma$, etc.
(for the detailed analysis of the constraints, see section V).

As an illustration, the mass spectra without and with the
constraints from the low energy experiments are given in Fig. 1
and 2, respectively, where (a) and (b) are for $A_0=0, -1000$GeV
respectively. Fig. 1 is plotted for $M_{1/2}=200 GeV,
\tan\beta=40$ and Fig. 2 for $M_{1/2}=500 GeV, \tan\beta=40$. One
can see from the {\bf Figs}. \ref{isa}, \ref{con} that the mass
spectrum lifts when $|A_0|$ decreases. When the constraints from
the low energy experiments are imposed, some mass spectra are
excluded and for the allowed spectra the masses of sparticles are
similar to those without the constraints.

\begin{figure}
{\includegraphics[width=5.7cm] {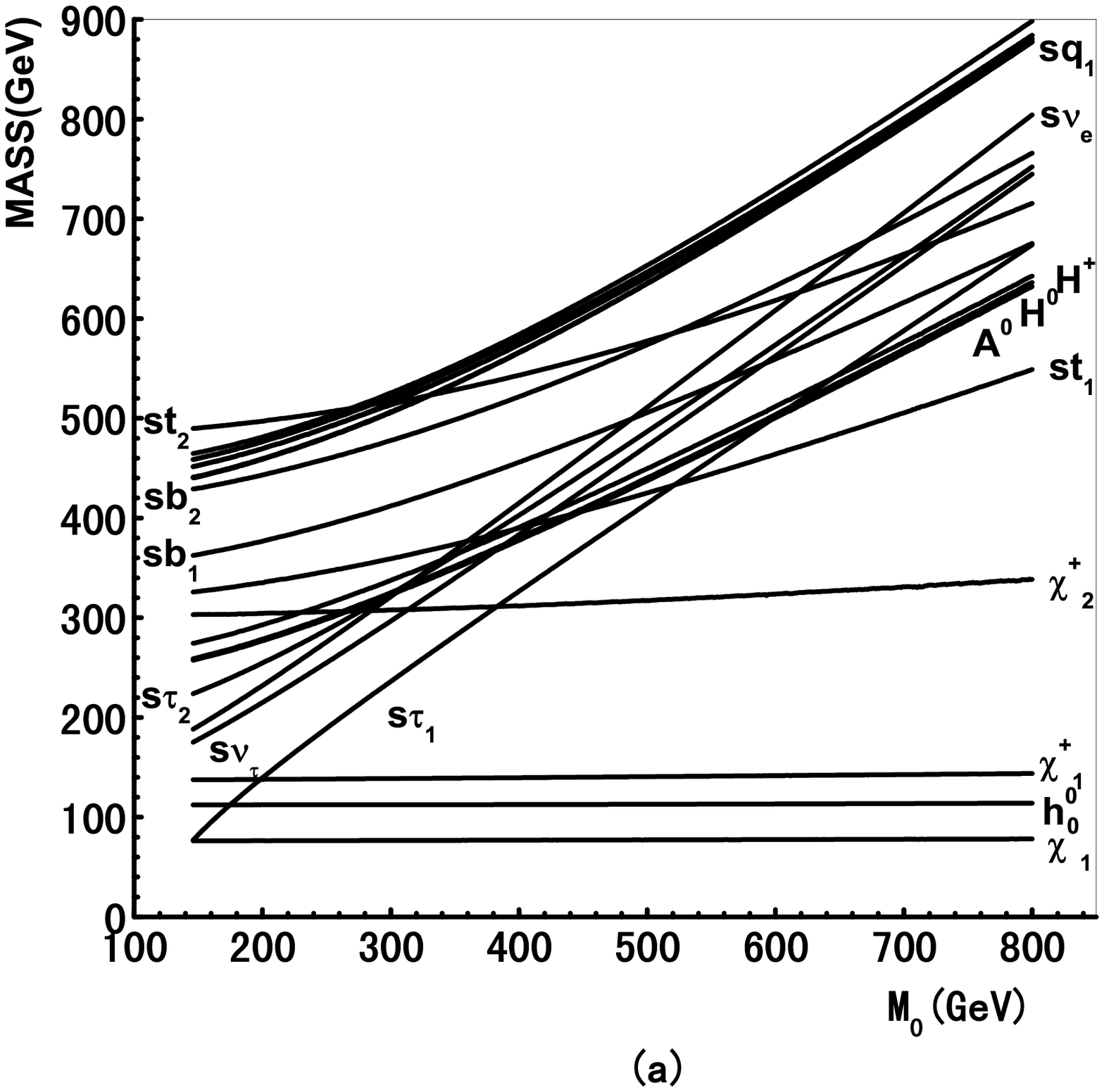}}
{\includegraphics[width=6cm] {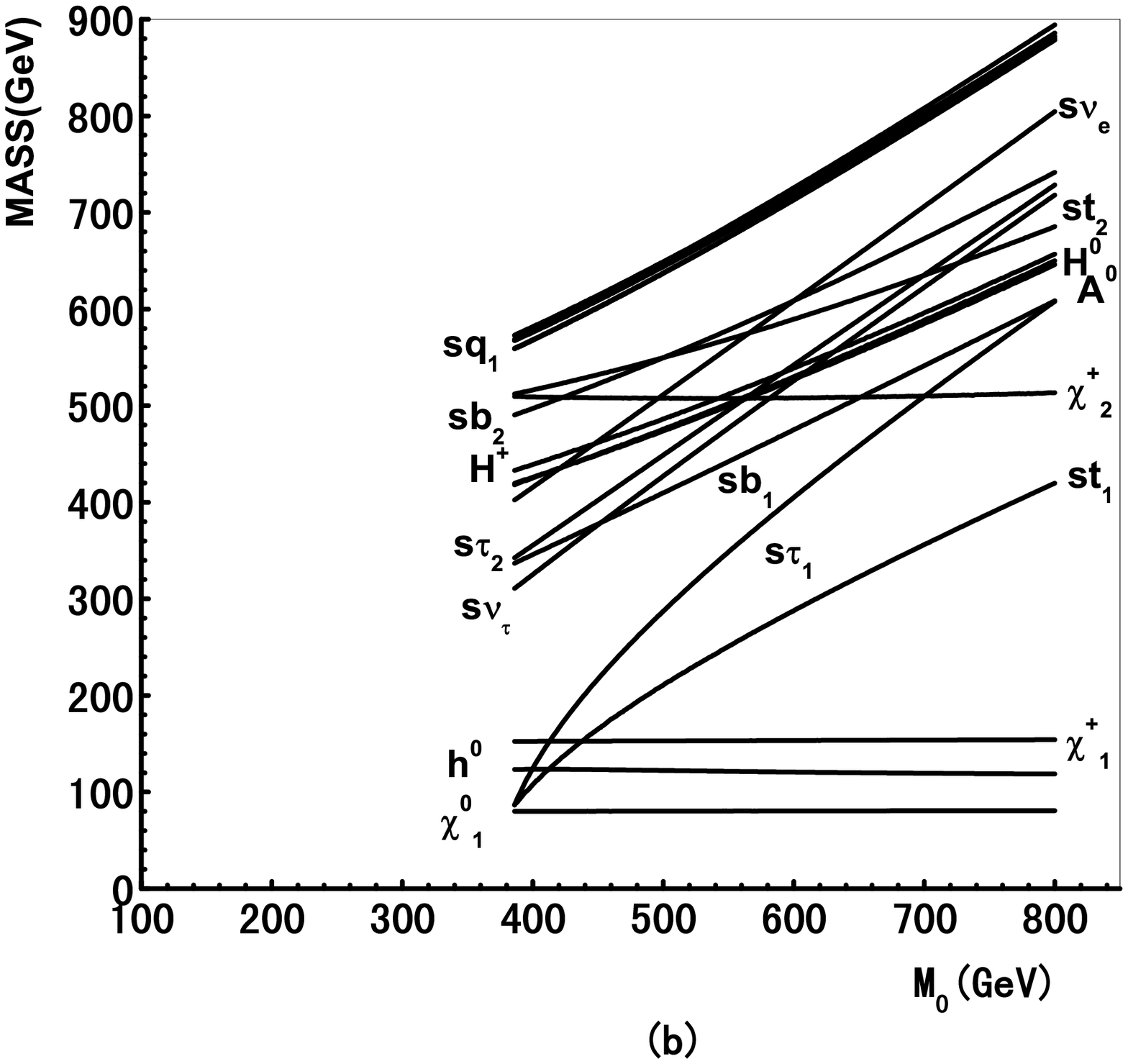}} \caption{ \label{isa}
The mass spectrum versus $m_0$ for fixed $M_{1/2}$=200GeV,
$\tan\beta$=40, $\delta^d_{23RR}=(0.12+0.03i)$, and sign($\mu$)=+1
without the constraints from the low energy experiments imposed.
(a) is for $A_0=0$. (b) is for $A_0$=-1000GeV.}
\end{figure}
\begin{figure}
{\includegraphics[width=6cm] {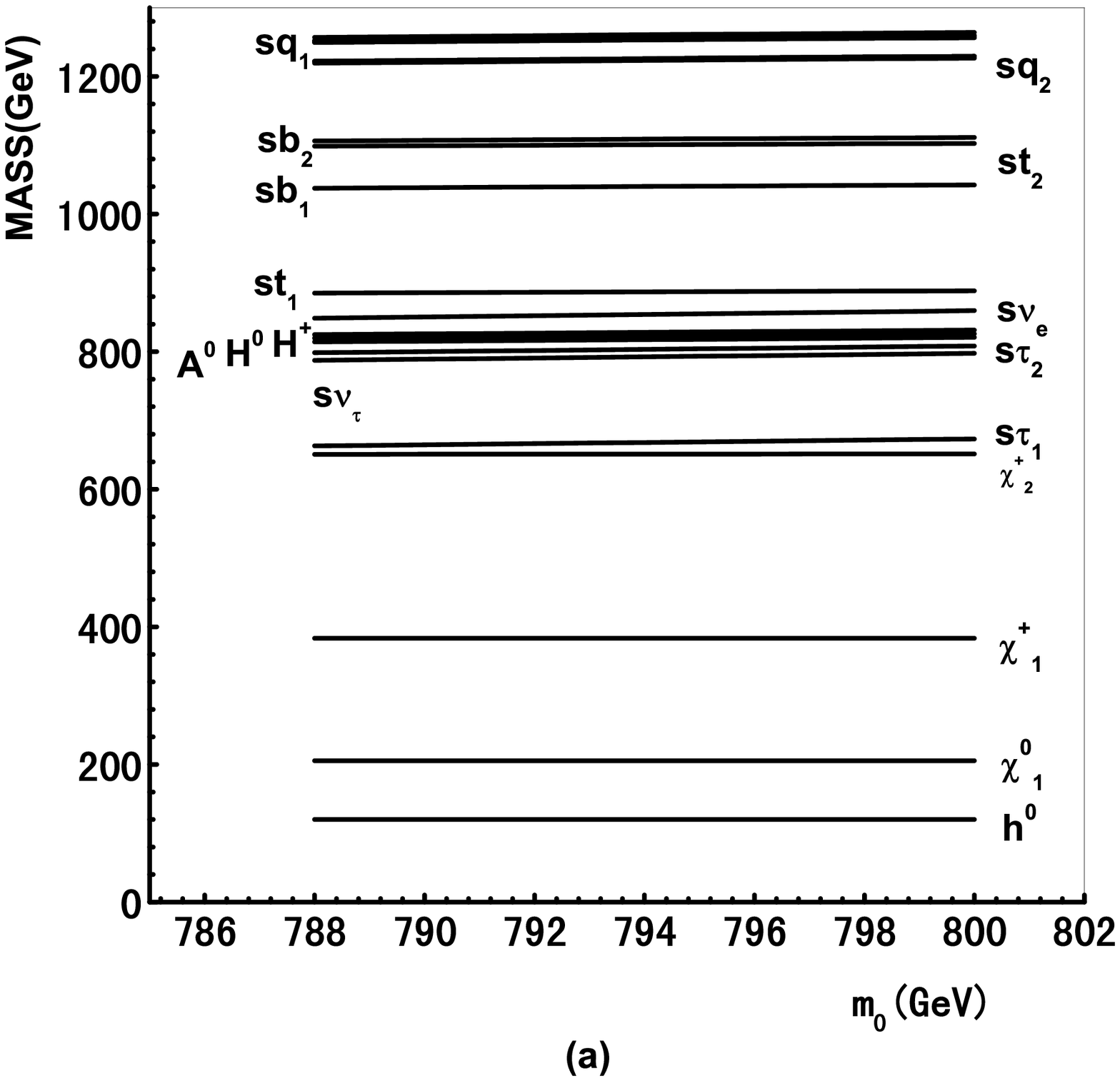}}
{\includegraphics[width=6cm] {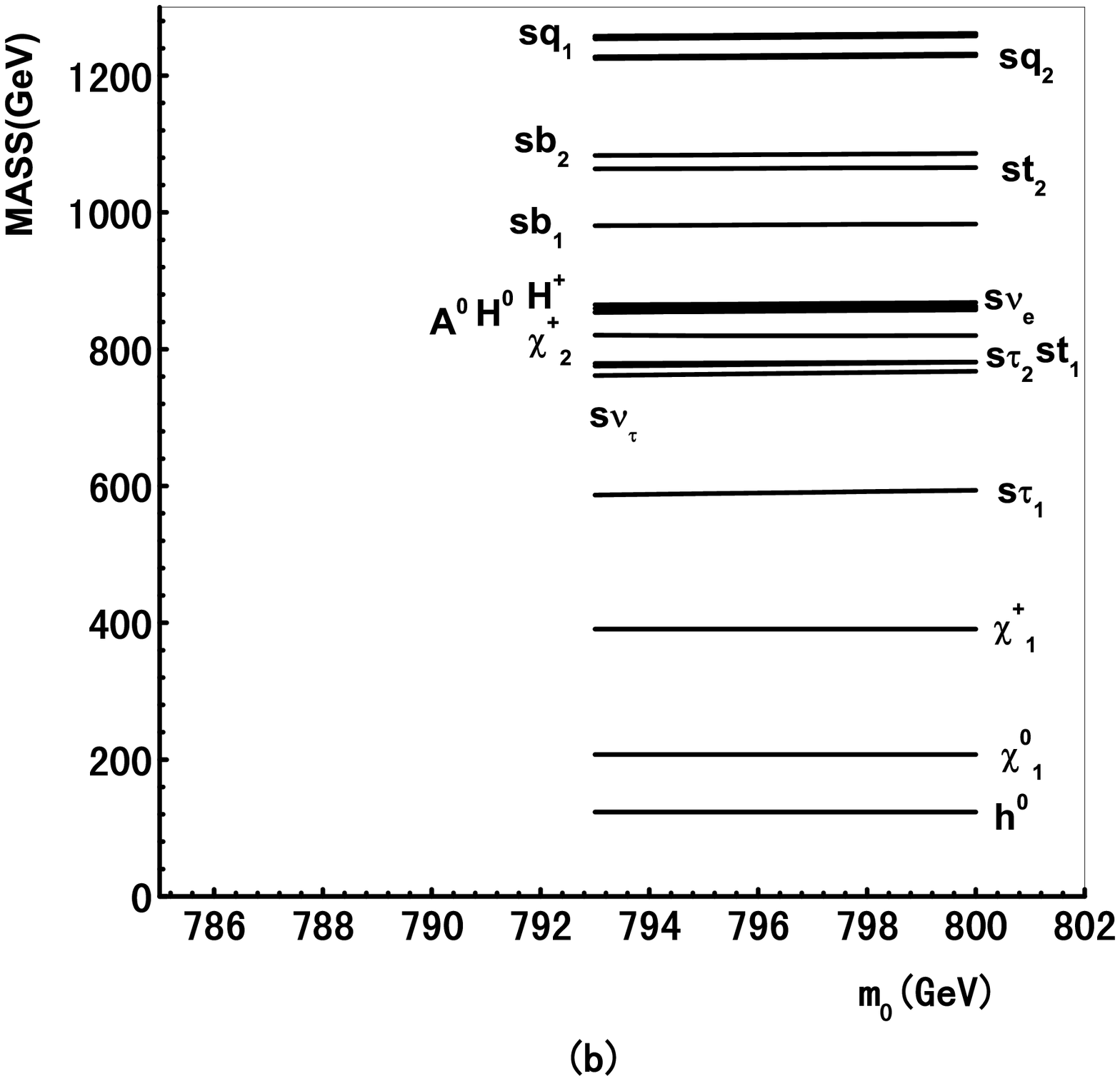}} \caption{ \label{con}
The mass spectrum versus $m_0$ for fixed $M_{1/2}$=500GeV,
$\tan\beta$=40, $\delta^d_{23RR}=(0.12+0.03i)$, and sign($\mu$)=+1
with the constraints from the low energy experiments imposed. (a)
is for $A_0=0$. (b) is for $A_0$=-1000GeV.}
\end{figure}

\section{Effective Hamiltonian for $b \rightarrow s$ transition}
The effective Hamiltonian for $b \rightarrow s$ transition can be
expressed as\cite{bur,chw}
\begin{eqnarray}\label{eff}
 {\cal H}_{\rm eff} &=& \frac{G_F}{\sqrt2} \sum_{p=u,c} \!
   V_{pb} V^*_{ps} \bigg(C_1\,Q_1^p + C_2\,Q_2^p
   + \!\sum_{i=3,\dots, 16}\![ C_i\,Q_i+ C_i^\prime\,Q_i^\prime]
   \nonumber \\&& + C_{7\gamma}\,Q_{7\gamma}
   + C_{8g}\,Q_{8g}
   + C_{7\gamma}^\prime\,Q_{7\gamma}^\prime
   + C_{8g}^\prime \,Q_{8g}^\prime \, \bigg) + \mbox{h.c.} \,
\end{eqnarray}
Here $Q_i$ are quark and gluon operators and are given by
\footnote{For the operators in SM we use the conventions in
Ref.\cite{bbns1} where $Q_1$ and $Q_2$ are exchanged each other
with respect to the convention in most of papers.}
\begin{eqnarray}
&&Q_1^p = (\bar s_\alpha p_\beta)_{V-A} (\bar p_\beta
b_\alpha)_{V-A},\hspace{2.3cm}
Q_2^p = (\bar s_\alpha p_\alpha)_{V-A} (\bar p_\beta b_\beta)_{V-A},\nonumber\\
&&Q_{3(5)} = (\bar s_\alpha b_\alpha)_{V-A}\sum_{q} (\bar q_\beta
q_\beta)_{V-(+)A},\hspace{1cm} Q_{4(6)} = (\bar s_\alpha
b_\beta)_{V-A}\sum_{q}
(\bar q_\beta q_\alpha)_{V-(+)A},\nonumber\\
&&Q_{7(9)} = {3\over 2}(\bar s_\alpha b_\alpha)_{V-A}\sum_{q}
e_{q}(\bar q_\beta q_\beta)_{V+(-)A},\hspace{0.4cm}Q_{8(10)} ={3\over 2}
(\bar s_\alpha b_\beta)_{V-A}\sum_{q}
e_{q}(\bar q_\beta q_\alpha)_{V+(-)A},\nonumber\\
&&Q_{11(13)} = (\bar s\, b)_{S+P} \sum_q\,{m_q\over m_b} (\bar q\,
q)_{S-(+)P}\,,\nnb\\&&  Q_{12(14)} = (\bar s_i \,b_j)_{S+P}
 \sum_q\,{m_q\over m_b}(\bar q_j \,q_i)_{S-(+)P} \,, \nonumber\\
&&Q_{15} = \bar s \,\sigma^{\mu\nu}(1+\gamma_5) \,b
\sum_q\,{m_q\over m_b}
    \bar q\, \sigma_{\mu\nu}(1+\gamma_5)\,q \,,\nnb\\&&
Q_{16} = \bar s_i \,\sigma^{\mu\nu}(1+\gamma_5) \,b_j \sum_q\,
    {m_q\over m_b} \bar q_j\, \sigma_{\mu\nu}(1+\gamma_5) \,q_i
    \, ,\nnb\\
&&Q_{7\gamma} = {e\over 8\pi^2} m_b \bar s_\alpha \sigma^{\mu\nu}
F_{\mu\nu}
(1+\gamma_5)b_\beta, \nonumber\\
&&Q_{8g} = {g_s\over 8\pi^2} m_b \bar s_\alpha \sigma^{\mu\nu}
G_{\mu\nu}^a {\lambda_a^{\alpha \beta}\over 2}(1+\gamma_5)b_\beta,
\end{eqnarray}
where $(\bar q_1 q_2)_{V\pm A} =\bar q_1\gamma^\mu(1\pm\gamma_5)
q_2$, $(\bar q_1 q_2)_{S\pm P}=\bar q_1(1\pm\gamma_5)q_2$
\footnote{Strictly speaking, the sum over q in expressions of
$Q_i$ (i=11,...,16) should be separated into two parts: one is for
q=u, c, i.e., upper type quarks, the other for q=d, s, b, i.e.,
down type quarks, because the couplings of upper type quarks to
NHBs are different from those of down type quarks. In the case of
large $\tan\beta$ the former is suppressed by $\tan^{-1}\beta$
with respect to the latter and consequently can be neglected.
Hereafter we use, e.g., $C_{11}^c$ to denote the Wilson
coefficient of the operator $Q_{11}= (\bar s\, b)_{S+P}
\,{m_c\over m_b} (\bar c\, c)_{S-P}$.}, $p=u, c$, $q = u,d,s,c,b$,
$e_{q}$ is the electric charge number of $q$ quark, $\lambda_a$ is
the color SU(3) Gell-Mann matrix, $\alpha$ and $\beta$ are color
indices, and $F_{\mu\nu}$ ($G_{\mu\nu}$) are the photon (gluon)
fields strength.

The primed operators, the counterpart of the unprimed operators,
are obtained by replacing the chiralities in the corresponding
unprimed operators with opposite ones. The SUSY contributions to
Wilson coefficients have been calculated by using the vertex
method in ref.~\cite{hw}. The SUSY contributions due to gluino box
and penguin diagrams to the relevant Wilson coefficients at the
$m_W$ scale in MIA with double insertions, as investigated in
ref~\cite{everett01}, which are non-negligible if the mixing
between left-handed and right-handed sbottoms is large have been
given in refs.\cite{hw,chw1}. We calculate the chargino
contributions in MIA with double insertions and results are

\begin{eqnarray}\label{wilson1}
C_3 &=& \frac{1}{48} \frac{\alpha_s}{4\pi}
 \frac{m^2_w}{m^2_{\tilde{\chi}^\pm_i}}
 [\frac{K^\ast_{22}}{K^\ast_{32}} V_{i1} V^\ast_{i1}
    (F_6(x) \delta^{uLL}_{23} +
    F^\prime_6(x) \delta^{uLR}_{23} \delta^{uLR}_{33} ) \nonumber\\
  &&- \frac{K^\ast_{22}}{K^\ast_{32}} V_{i1} V^\ast_{i2} h_t
    (F_6(x) \delta^{uLR}_{23} +
    F^\prime_6(x) \delta^{uLL}_{23} \delta^{uLR}_{33} ) \nonumber\\
  &&- 2 V_{i1} V^\ast_{i2} h_t F_6(x) \delta^{uLR}_{33} ] \nonumber\\
C_4 &=& - \frac{1}{144} \frac{\alpha_s}{4\pi}
 \frac{m^2_w}{m^2_{\tilde{\chi}^\pm_i}}
 [\frac{K^\ast_{22}}{K^\ast_{32}} V_{i1} V^\ast_{i1}
    (F_6(x) \delta^{uLL}_{23} +
    F^\prime_6(x) \delta^{uLR}_{23} \delta^{uLR}_{33} ) \nonumber\\
  &&- \frac{K^\ast_{22}}{K^\ast_{32}} V_{i1} V^\ast_{i2} h_t
    (F_6(x) \delta^{uLR}_{23} +
    F^\prime_6(x) \delta^{uLL}_{23} \delta^{uLR}_{33} ) \nonumber\\
  &&- 2 V_{i1} V^\ast_{i2} h_t F_6(x) \delta^{uLR}_{33} ] \nonumber\\
C_5 &=& C_3 \nonumber\\
C_6 &=& C_4 \nonumber\\
C_{7\gamma} &=& \frac{1}{72} \frac{m^2_w}{m^2_{\tilde{\chi}^\pm_i}}
  [\frac{K^\ast_{22}}{K^\ast_{32}} V_{i1} V^\ast_{i1}
    (F_{12}(x) \delta^{uLL}_{23} +
    F^\prime_{12}(x) \delta^{uLR}_{23} \delta^{uLR}_{33} ) \nonumber\\
  &&- \frac{K^\ast_{22}}{K^\ast_{32}} V_{i1} V^\ast_{i2} h_t
    (F_{12}(x) \delta^{uLR}_{23} +
    F^\prime_{12}(x) \delta^{uLL}_{23} \delta^{uLR}_{33} ) \nonumber\\
  &&- 2 V_{i1} V^\ast_{i2} h_t F_{12}(x) \delta^{uLR}_{33} ] \nonumber\\
 &&+ \frac{1}{9} \frac{m^2_w}{m_b m_{\tilde{\chi}^\pm_i}}
  [ - \frac{K^\ast_{22}}{K^\ast_{32}} V_{i1} U^\ast_{i2} h_b
    (F_{34}(x) \delta^{uLL}_{23} +
    F^\prime_{34}(x) \delta^{uLR}_{23} \delta^{uLR}_{33} ) \nonumber\\
  &&+ V_{i2} U^\ast_{i2} h_t h_b F_{34}(x) \delta^{uLR}_{33} ] \nonumber\\
C_{8g} &=& \frac{1}{24} \frac{m^2_w}{m^2_{\tilde{\chi}^\pm_i}}
  [\frac{K^\ast_{22}}{K^\ast_{32}} V_{i1} V^\ast_{i1}
    (F_2(x) \delta^{uLL}_{23} +
    F^\prime_2(x) \delta^{uLR}_{23} \delta^{uLR}_{33} ) \nonumber\\
  &&- \frac{K^\ast_{22}}{K^\ast_{32}} V_{i1} V^\ast_{i2} h_t
    (F_2(x) \delta^{uLR}_{23} +
    F^\prime_2(x) \delta^{uLL}_{23} \delta^{uLR}_{33} ) \nonumber\\
  &&- 2 V_{i1} V^\ast_{i2} h_t F_2(x) \delta^{uLR}_{33} ] \nonumber\\
 &&+ \frac{1}{6} \frac{m^2_w}{m_b m_{\tilde{\chi}^\pm_i}}
  [ - \frac{K^\ast_{22}}{K^\ast_{32}} V_{i1} U^\ast_{i2} h_b
    (F_4(x) \delta^{uLL}_{23} +
    F^\prime_4(x) \delta^{uLR}_{23} \delta^{uLR}_{33} ) \nonumber\\
  &&+ V_{i2} U^\ast_{i2} h_t h_b F_4(x) \delta^{uLR}_{33} ] \nonumber\\
C_{11}^{(\prime)} &=& \frac{e^2}{16\pi^2} \frac{m_b}{m_l}
 [C_{Q_1}^{(\prime)} \mp C_{Q_2}^{(\prime)}] \nonumber\\
C_{13}^{(\prime)} &=& \frac{e^2}{16\pi^2} \frac{m_b}{m_l}
 [C_{Q_1}^{(\prime)} \pm C_{Q_2}^{(\prime)}],\nnb\\
 C_i^{(\prime)} &=& 0,\,\,\,\, i=12,14,15,16
\end{eqnarray}
with $C_{Q_{1,2}}^{(\prime)}$\footnote{The operator of
$Q^{(\prime)}_{1,2}$ is defined as
$Q_1 = \frac{e^2}{8\pi^2} [\bar{s} (1 + \gamma_5) b][\bar{l} l]$,
$Q_1^\prime = \frac{e^2}{8\pi^2} [\bar{s} (1 - \gamma_5) b][\bar{l} l]$,
$Q_2 = \frac{e^2}{8\pi^2} [\bar{s} (1 + \gamma_5) b][\bar{l} \gamma_5 l]$,
$Q_2^\prime = \frac{e^2}{8\pi^2} [\bar{s} (1 - \gamma_5) b][\bar{l}
\gamma_5 l]$ } as
\begin{eqnarray}\label{wilson2}
C_{Q_1} &=& - \frac{1}{2 s^2_w} \frac{m_l m_{\tilde{\chi}^\pm_i}}{m^2_{H^0}}
  \frac{c^2_\alpha + r_s s^2_\alpha}{c^2_\beta}
  [- \frac{K^\ast_{22}}{K^\ast_{32}} V_{i1} U^\ast_{i2} h_b
    (F_{b0}(x) \delta^{uLL}_{23} +
    F^\prime_{b0}(x) \delta^{uLR}_{23} \delta^{uLR}_{33} ) \nonumber\\
  &&+ V_{i2} U^\ast_{i2} h_t h_b F_{b0}(x) \delta^{uLR}_{33} ] \nonumber\\
C_{Q_2} &=& \frac{1}{2 s^2_w} \frac{m_l m_{\tilde{\chi}^\pm_i}}{m^2_{H^0}}
  (r_p + \tan^2\beta)
  [- \frac{K^\ast_{22}}{K^\ast_{32}} V_{i1} U^\ast_{i2} h_b
    (F_{b0}(x) \delta^{uLL}_{23} +
    F^\prime_{b0}(x) \delta^{uLR}_{23} \delta^{uLR}_{33} ) \nonumber\\
  &&+ V_{i2} U^\ast_{i2} h_t h_b F_{b0}(x) \delta^{uLR}_{33} ]
\end{eqnarray}
where $h_t = \frac{m_t}{\sqrt{2} m_w s_\beta}$, $h_b =
\frac{m_b}{\sqrt{2} m_w c_\beta}$, $r_s =
\frac{m^2_{H^0}}{m^2_{h^0}}$, $r_p = \frac{m^2_{A^0}}{m^2_{Z^0}}$,
and $x = m_{\tilde{q}}^2/m_{\tilde{\chi}^\pm_i}^2$ with
$m_{\tilde{q}}$ and $m_{\tilde{\chi}^\pm_i}$ being the common
squark mass and chargino masses respectively. The repeating
indices $i$ should sum over from $1$ to $2$. The one-loop
functions in Eq. (\ref{wilson1}) and (\ref{wilson2}) are given in
Appendix A. We have checked that the Wilson coefficient
$C_{7\gamma}$ with single insertion is the same as that given in
Ref.\cite{lunghi2000}. Differed from the single insertion results,
the LR or RL insertion also generates the QCD penguin operators
when one includes the double insertions.

For the processes we are interested in this paper, the Wilson
coefficients should run to the scale of $O(m_b)$. $C_1-C_{10}$ are
expanded to $O(\alpha_s)$ and NLO renormalization group equations
(RGEs) should be used. However for the $C_{8g}$ and $C_{7\gamma}$,
LO results should be sufficient. The details of the running of
these Wilson coefficients can be found in Ref. \cite{bur}. The one
loop anomalous dimension matrices of the NHB induced operators can
be divided into two distangled groups~\cite{adm}
\begin{eqnarray}
\gamma^{(RL)}=\begin{tabular}{c|cccc} &$Q_{11}$&$Q_{12}$\\\hline
$Q_{11}$&$-16$&0\\
$Q_{12}$&-6&$2$
\end{tabular}
\end{eqnarray}
and
\begin{eqnarray}
\gamma^{(RR)}=\begin{tabular}{c|cccc}
&$Q_{13}$&$Q_{14}$&$Q_{15}$&$Q_{16}$\\\hline
$Q_{13}$&$-16$&0&1/3&$-1$\\
$Q_{14}$&$-6$&$2$&$-1$/2&$-7$/6\\
$Q_{15}$&16&$-4$8&$16/3$&0\\
$Q_{16}$&$-24$&$-56$&6&$-38/3$
\end{tabular}
\end{eqnarray}
Here and hereafter the factor, $\frac{\alpha_s}{4\pi}$, is
suppressed (i.e., the anomalous dimension matrix for $Q_{11,12}$
is $\frac{\alpha_s}{4\pi}\,\,\gamma^{(RL)}$, etc.). For
$Q_i^\prime$ operators we have
\begin{eqnarray}
\gamma^{(LR)}=\gamma^{(RL)}~~~~~{\rm and}~~~~~
\gamma^{(LL)}=\gamma^{(RR)}\,.
\end{eqnarray}
Because at present no NLO Wilson coefficients $C_i^{(\prime)}$,
i=11,...,16, are available we use the LO running of them in the
paper.

There is the mixing of the new operators induced by NHBs with the
operators in SM. The leading order anomalous dimensions have been
given in Refs.\cite{bghw,hk}. We list those relevant to our
calculations in the following. Defining \begin{eqnarray} \label{oq}
O_i=\frac{g^2}{16\pi^2}Q_{12+i},\,\,\,\,i=1,2,3,4,\end{eqnarray} one has
\begin{eqnarray} \gamma^{(RD)}=\begin{tabular}{c|cccc}
&$Q_{7\gamma}$&$Q_{8g}$\\\hline
$O_{1}$&$-1/3$&1\\
$O_{2}$&$-1$&$0$\\
$O_{3}$&28/3&$-4$\\
$O_{4}$&20/3&$-8$
\end{tabular}\label{q8g}
\end{eqnarray} The mixing of $Q_{11,12}$ onto the QCD penguin operators is
\begin{eqnarray} \gamma^{(MQ)}=\begin{tabular}{c|cccc}
&$Q_{3}$&$Q_{4}$&$Q_5$&$Q_6$\\\hline
$O_{11}$&$1/9$&$-1/3$&1/9&$-1/3$\\
$O_{12}$&0&$0$&0&0
\end{tabular}
\end{eqnarray} For the mixing among the primmed operators, we have
\begin{eqnarray}
\gamma^{(LD^\prime)}=\gamma^{(RD)}~~~~~{\rm and}~~~~~ \gamma^{(M^\prime
Q^\prime)}=\gamma^{(MQ)}\,.
\end{eqnarray}
The mixing of the new operators induced by NHBs with the operators
in SM has non-negligible effects on the Wilson coefficients of the
SM operators at the $O(m_b)$ scale. In particular, the Wilson
coefficient of the chromo-magnetic dipole operator $C_{8g}$ at the
$O(m_b)$ scale, which has a large effect to $S_{M K}$ ($M=\phi,
\eta^\prime$), can significantly enhance due to the mixing. To see
it explicitly we concentrate on the mixing of $O_i$ (for its
definition, see Eq.(\ref{oq})) onto $Q_{8g}$. Solving RGEs, we
have
\begin{eqnarray}
C_{8g}(\mu)&=&\sum_{c=1,...,4}A(\mu_0)(\eta(\mu)^{\gamma_{cc}/2\beta_0}-
\eta(\mu)^{\gamma_{8g8g}/2\beta_0})+
C_{8g}(\mu_0)\eta^{\gamma_{8g8g}
/2\beta_0},\label{c8g}\\
A(\mu_0)&=&\sum_{a,b=1,...,4}\gamma_{a1}
V^{-1}_{ac}V_{cb}C_b(\mu_0)/(\gamma_{cc}
-\gamma_{8g8g})\\
\eta&=&\alpha_s(\mu_0)/\alpha_s(\mu), \end{eqnarray} where $V$ and
$\gamma_{aa}$ are given by \begin{eqnarray} V(\gamma^{(RR)}+
2\beta_0\,I)\,V^{-1}={\rm diag}(\gamma_{11},\gamma_{22},
\gamma_{33},\gamma_{44}).\end{eqnarray} with $I$ being the $4\times 4$ unit
matrix. Using \begin{eqnarray} C_a(\mu_0)&=& C_1(\mu_0)\delta_{a1}\end{eqnarray} and Eq.
(\ref{q8g}), Eq. (\ref{c8g}) reduces to
\begin{eqnarray} C_{8g}(\mu)=0.68
C_{8g}(\mu_0) - 3.2 C_{13}(\mu_0),\end{eqnarray}
where $C_1(\mu_0)=\frac{4\pi}{\alpha_s}C_{13}(\mu_0)$ has been used.

In our numerical calculations we neglect
the contributions of \rm EW penguin operators $Q_{7,...10}$
since they are small compared with those of other operators.

The hadronic matrix elements of operators have been calculated in
BBNS approach in refs.\cite{bbns1,chw,chw1}. We use the results in
ref.\cite{chw1}. The effective Hamiltonian (\ref{eff}) results the
following decay amplitude for $B_{d}^{0} \rightarrow \phi K_S$.
\bea &&A(B\to \phi K_S) = {G_F\over \sqrt{2}} A ,\nnb \\\label{ap}
&& A= A^o+A^{o^\prime}+A^n,
 \eea where we have divided the decay amplitude into
three parts in order to see explicitly the effects of new
operators in the SUSY SO(10). One, $A^0$, has the same form as
that in SM, the second, $A^{o^\prime}$ is for primmed counterparts
of the SM operators, and the third, $A^n$, is new which comes from
the contributions of Higgs penguin induced operators. In
eq.(\ref{ap}), \bea
 A^o&=&\langle \phi|\bar s\gamma_\mu s |0\rangle  \langle K|\bar
s \gamma^\mu b|B\rangle \times\sum_{p=u,c}  V_{pb} V^*_{ps}\left[
a_3 +a_4^p+a_5 - {1\over 2}(a_7 + a_9 + a_{10}^p )
 \right],  \label{ao} \\
 A^{o^\prime}&=&A^o(C_i\to C_i^\prime),\eea where $a_i$'s have been given in
Refs.\cite{hz,hmw}. The hadronic matrix element of the vector
current can be parameterized as $\langle K|\bar s \gamma^\mu
b|B\rangle  = F_1^{B\to K}(q^2) (p_B^\mu + p_K^\mu) +(F^{B\to
K}_0(q^2)-F^{B\to K}_1(q^2)) (m_B^2-m_K^2)q^\mu/q^2$. For the
matrix element of the vector current between the vacuum and
$\phi$, we have $\langle \phi| \bar s \gamma_\mu b | 0 \rangle =
m_\phi f_\phi \epsilon^\phi_\mu$.

 $A^n$ in eq.({\ref{ap}), to the $\alpha_s$ order, in the heavy quark
limit is given as\cite{chw} \bea
A^n&=&A^n(C_i)+A^n(C_i\rightarrow C_i^\prime),\nnb\\
A^n(C_i)&=&\langle \phi|\bar s\gamma_\mu s |0\rangle \langle
K|\bar s \gamma^\mu b|B\rangle \,(- V_{tb} V^*_{ts}) \left[
a_4^{neu}+ {m_s \over m_b}\left( -{1\over 2}a_{12}+ {4 m_s\over
m_b}\,a_{15} \right) \right] \,.\label{an}\end{eqnarray} where
$a_i$'s have been given in Refs.\cite{chw,chw1}.

The time-dependent $CP$-asymmetry $S_{\phi K}$ is given by \bea
S_{\phi K} = \frac{2\,\mathrm{Im} \lambda_{\phi
K}}{1+|\lambda_{\phi K}|^2} \; .\label{defi} \eea Here
$\lambda_{\phi K}$ is defined as
\begin{eqnarray}
\lambda_{\phi k} &=& \left(\frac{q}{p}\right)_B
\frac{\mathcal{A}(\overline{B} \rightarrow \phi
K_S)}{\mathcal{A}(B \rightarrow \phi K_S)}.
\end{eqnarray}

\section{Numerical Results}
\subsection{Parameters input}
In our numerical calculations the following values are needed:
\begin{itemize}
 \item{\bf Lifetime, mass and decay constants}
\begin{eqnarray}
\begin{array}{ccccc}
&\tau(B^0)=1.56\times 10^{-12}s,&
 M_B = 5.28 {\rm GeV}, &m_b =4.2{\rm GeV},&\\
 &m_c =1.3 {\rm GeV},& m_s=100{\rm MeV},

 &f_{\phi} = 0.237 {\rm GeV},

\end{array}
\end{eqnarray}
\item{\bf Wolfenstein parameters}\\
We use the Wolfenstein parameters fitted by Ciuchini et al\cite{ciuchini}:
\begin{eqnarray}
\begin{array}{ll@{\,}@{\qquad}l}
&A=0.819\pm 0.040&\lambda=0.2237\pm 0.0033\mbox{,}
\\
&{\bar \rho}=\rho (1-\lambda^2/2) = 0.224\pm 0.038\mbox{,}&\quad \rho
= 0.230\pm 0.039\mbox{,}\\
& {\bar \eta}=\eta (1-\lambda^2/2)
= 0.317\pm 0.040\mbox{,}&\eta=0.325\pm 0.039\mbox{,}\\
&\gamma = (54.8\pm 6.2)^\circ\mbox{,}& \sqrt{\rho^2+\eta^2} = 0.398\pm
0.040\,.
\end{array}
\end{eqnarray}
\item{\bf Form factors}\\
In the paper we need the form factors: $F^{B\to K}(0)=0.34$.
\end{itemize}

\subsection{Constraints from experiments}
We impose two important constraints from $B\to X_s \gamma$ and
$B_s\to \mu^+\mu^-$. Considering the theoretical uncertainties, we
take $2.0\times 10^{-4} < {\rm Br}(B\to X_s \gamma)< 4.5\times
10^{-4}$, as generally analyzed in literatures.
Phenomenologically, Br($B\to X_s \gamma$) directly constrains
$|C_{7\gamma}(m_b)|^2 + |C^\prime_{7\gamma}(m_b)|^2$ at the
leading order. Due to the strong enhancement factor
$m_{\tilde{g}}/m_b$ associated with single $\delta^{dLR(RL)}_{23}$
insertion term in $C^{(\prime)}_{7\gamma}(m_b)$,
$\delta^{dLR(RL)}_{23}$ ($\sim 10^{-2}$) are more severely
constrained than $\delta^{dLL(RR)}_{23}$. However, if the
left-right mixing of scalar bottom quark $\delta^{dLR(RL)}_{33}$ is
large ($\sim 0.5$), $\delta^{dLL(RR)}_{23}$ is constrained to be
order of $10^{-2}$ since the double insertion term
$\delta^{dLL(RR)}_{23} \delta^{dLR(LR*)}_{33}$ is also enhanced by
$m_{\tilde{g}}/m_b$. Nevertheless, in the large $\tan\beta$ case
the chargino contribution can destructively interfere with the SM
(plus the charged Higgs) contribution so that the constraint can
be easily satisfied. The branching ratio $B_s \rightarrow \mu^+
\mu^-$ in SUSY models is given as
\begin{eqnarray}\label{bsmu}
{\rm Br}(B_s \rightarrow \mu^+ \mu^-) &=& \frac{G_F^2
\alpha^2_{\rm em}}{64 \pi^3} m^3_{B_s} \tau_{B_s} f^2_{B_s}
|\lambda_t|^2 \sqrt{1 - 4 \widehat{m}^2}
[(1 - 4\widehat{m}^2) |C_{Q_1}(m_b) - C^\prime_{Q_1}(m_b)|^2 + \nonumber\\
&& |C_{Q_2}(m_b) - C^\prime_{Q_2}(m_b) + 2\widehat{m}(C_{10}(m_b)
- C^\prime_{10}(m_b) )|^2]
\end{eqnarray}
where $\widehat{m} = m_\mu/m_{B_s}$. In the middle and large
$\tan\beta$ case the term proportional to $(C_{10}-C_{10}^\prime)$
in Eq. (\ref{bsmu}) can be neglected. The new CDF experimental
upper bound of ${\rm Br}(B_s\to \mu^+\mu^-)$ is $5.8\times
10^{-7}$~\cite{bsmu} at $90\%$ confidence level. To translate it
into the constraint on $C_{Q_{11,13}}^{(\prime)}$, we have
\begin{eqnarray}\label{bsmumu}
\sqrt{|C_{Q_{11}}(m_W)-C_{Q_{11}}^\prime(m_W)|^2 +
|C_{Q_{13}}(m_W)-C_{Q_{13}}^\prime(m_W)|^2}\lsim 0.047
\end{eqnarray} Because the bound constrains
$|C_{Q_i}-C_{Q_i}^\prime|$ (i=1, 2),
\footnote{$C_{Q_{1,2}}^{(\prime)}$ are the Wilson coefficients of
the operators $Q_{1,2}^{(\prime)}$ which are Higgs penguin induced
in leptonic and semileptonic B decays and their definition can be
found in Ref.~\cite{hy}. By substituting the quark-Higgs vertex
for the lepton-Higgs vertex it is straightforward to obtain Wilson
coefficients relevant to hadronic B decays.} and there is a flavor
non-diagonal element in the right-hand down-type squark mass
matrix at the high scale in the SUSY SO(10) model, we could have
values of $|C_{Q_i}|$ and $|C_{Q_i}^\prime|$ larger than those in
constrained MSSM (CMSSM) with universal boundary conditions at the
high scale and scenarios of the extended minimal flavor violation
in MSSM~\cite{kane} in which $|C_{Q_i}^\prime|$ is much smaller
than $|C_{Q_i}|$. However, since the flavor non-diagonal element
in the right-hand down-type squark mass matrix at the high scale
has almost no impact on the decoupled limit of Higgs boson masses
(i.e., $m_{H^0} \simeq m_{A^0} \simeq m_{H^\pm}$, $\sin(\alpha -
\beta) \simeq 1$), the decoupled Higgs sector in CMSSM remains in
our model unless the non universal conditions at the high scale
are assumed~\cite{chlw}. Therefore, $C_{Q_2}^{(\prime)}\approx \mp
C_{Q_1}^{(\prime)}$~\cite{hw} in our model, which leads to
$C_{Q_{13}}^{(\prime)}$ being almost zero, and consequently eq.
(\ref{bsmumu}) give a stringent constraint on
$C_{Q_{11}}^{(\prime)}$. At the same time we require that
predicted Br of $B\to X_s \mu^+\mu^-$ falls within 1 $\sigma$
experimental bounds.

We also impose the current experimental lower bound $\Delta M_s >
14.4 ps^{-1}$~\cite{msd} and experimental upper bound ${\rm Br}
(B\to X_s g)< 9\%$~\cite{bsg}. Because $\delta^{dLR(RL)}_{23}$ is
constrained to be order of $10^{-2}$ by Br($B \to X_s \gamma$),
their contribution to $\Delta M_s$ is small. The dominant
contribution to $\Delta M_s$ comes from $\delta^{dLL(RR)}_{23}$
insertion with both constructive and destructive effects compared
with the SM contribution, where the too large destructive effect
is ruled out, because SM prediction is only slightly above the
present experiment lower bound.

As pointed out in section III, due to the gluino-sbottom loop
diagram contribution and the mixing of NHB induced operators onto
the chromomagnetic dipole operator, the Wilson coefficients
$C_{8g}^{(\prime)}$ can be large,
which might lead to a too large Br of $B\to X_s g$.
So we need to impose the constraint from
experimental upper bound ${\rm Br} (B\to X_s g)< 9\%$.
A numerical analysis for $C_{8g}^\prime$=0 has been performed in Ref.\cite{hk}.
We carry out a similar analysis by setting both
$C_{8g}$ and $C_{8g}^\prime$ non-zero.

Furthermore, as analyzed in ref.~\cite{cmsvv},
there is the correlation between flavor changing squark
and slepton mass insertions in SUSY GUTs.
The correlation leads to a bound on $\delta_{23}^{dRR}$
from the rare decay $\tau\to\mu\gamma$.
We update the analyses with latest BELLE upper bound of
Br($\tau\to\mu\gamma$)$ < 3.1 \times 10^{-7}$~\cite{taumugamma}
at $90\%$ confidence level.

\subsection{Numerical results}
In numerical analysis we fix $\tan\beta=40$, sign($\mu$)=+1, and
$A_0=0, -1000$GeV. We scan $m_0, M_{1/2}$ in the range from
$100$GeV to $800$GeV by running ISAJET. Using the sparticle mass
spectrum and mixings at the EW scale, we calculate Wilson
coefficients and consequently Br and the time dependent CP
asymmetry in $B\to \phi K_S$ under additional constraints of
$\delta_{23}^{dRR}$ from Br($\tau\to\mu\gamma$), which phase
varies from 0 to $2\pi$.

Numerical results of the correlation between $S_{\phi K_S}$
and Br($B\to \phi K_S$) for $A_0=0$, $-1000$GeV are shown
in Figs.~\ref{figa0} and \ref{figa0m1000} respectively,
where (a) is for the SUSY contributions with only
gluino propagated in the loop,
(b) is for the all contributions (i.e., $W^\pm, H^\pm$, chargino,
gluino, neutrilino propagated in the loop\footnote{We neglect the
neutrilino contributions in numerical calculations because they
are small compared with other contributions.} ) included. Current
$1\sigma$ bounds are shown by the dashed lines. The case (a)
describes the direct consequence of existing a
$\delta_{23}^{dRR}$. Before we discuss the numerical results in
detail, a remark is in place. It is shown that in MSSM NHB
contributions can be significant with all relevant experimental
constraints imposed~\cite{chw}. However, NHB contributions are
limited to be very small in our case. In case of only SM and NHB
contributions included, both $S_{\phi K_S}$ and Br($B\to \phi
K_S$) almost do not change, compared with the SM, because the
Higgs sector is nearly decoupled with $m_{H^0} \simeq m_{A^0}
\simeq m_{H^\pm}$, $\sin(\alpha - \beta) \simeq 1$, as pointed out
above, in the regions of the parameter space which we take in the
SO(10) model (or other constrained MSSM), which leads to that
$C^{(\prime)}_{13}$ are almost equal to $0$ and consequently
$C^{(\prime)}_{11}$ are very small due to the constraint from
$B_s\to \mu^+\mu^-$.

\begin{figure}
{\includegraphics[width=6cm] {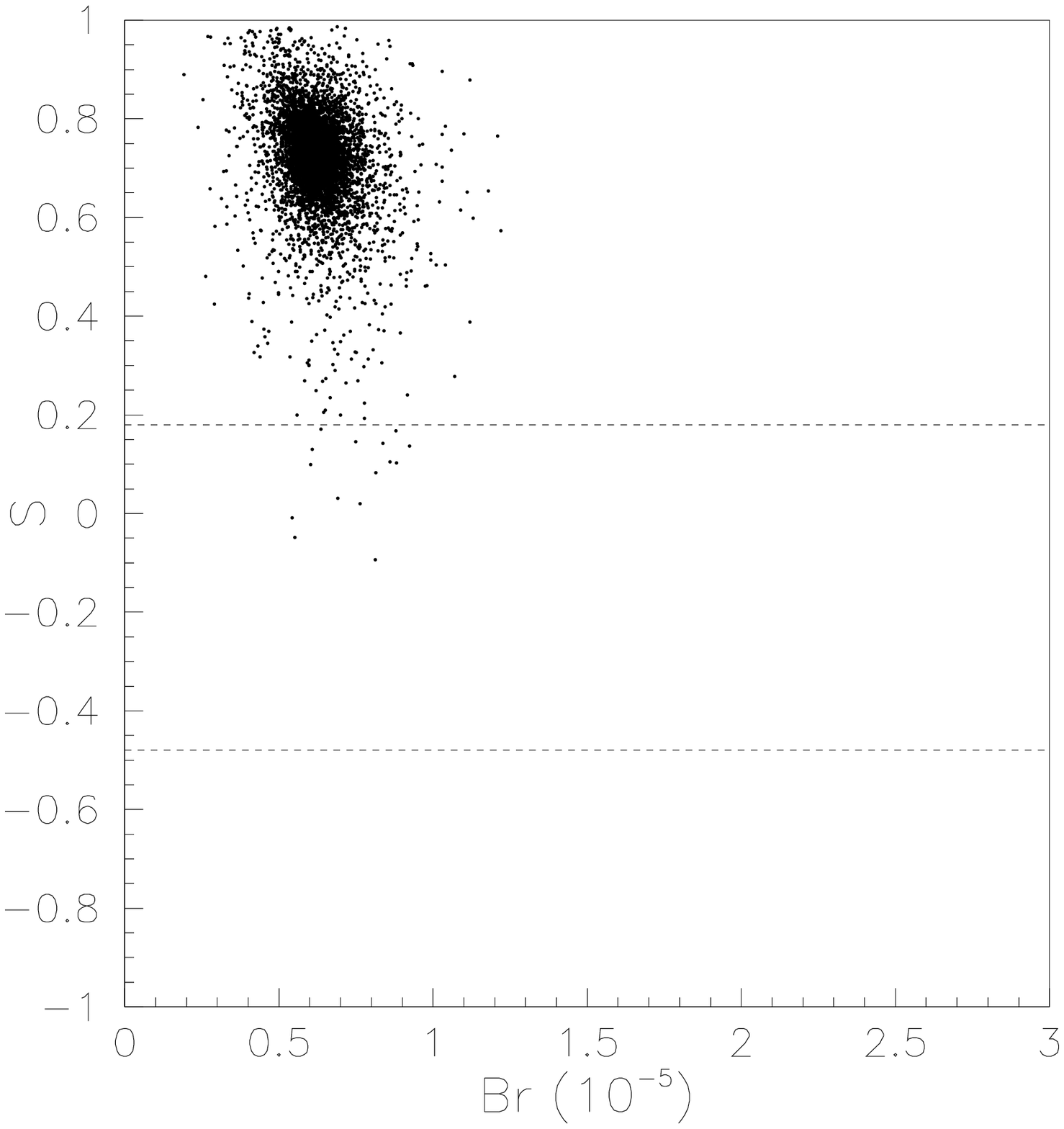}}
{\includegraphics[width=6cm] {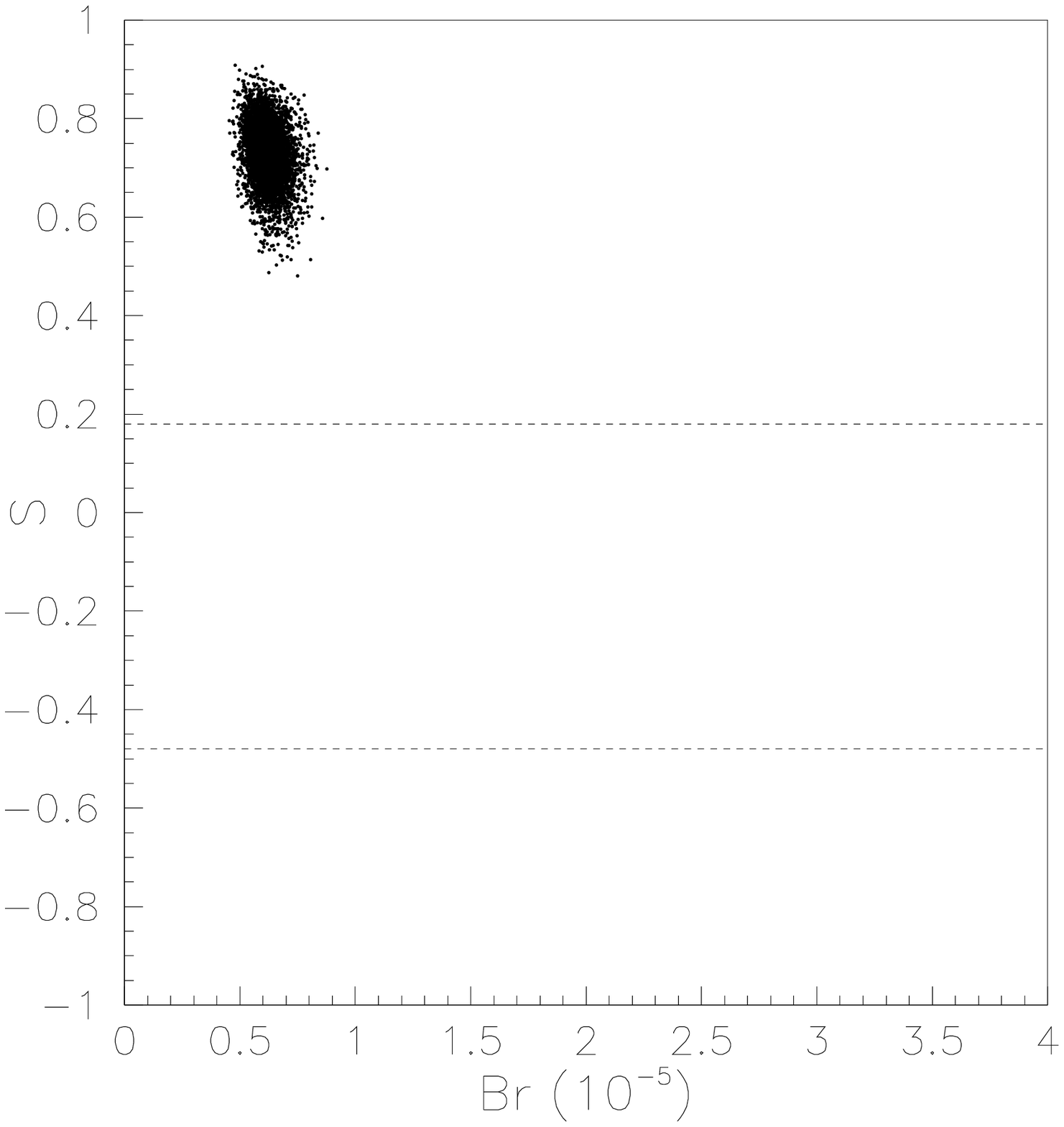}} \\
(a) \hspace{5cm} (b)
\caption{ \label{figa0}
The correlation between $S_{\phi K_S}$ and
Br($B\to \phi K_S$) for $A_0=0$.
(a) is for the SUSY contributions
with only gluino propagated in the loop,
(b) is for the all contributions included.
Current $1\sigma$ bounds are shown by the dashed lines.
}
\end{figure}

\begin{figure}
{\includegraphics[width=6cm] {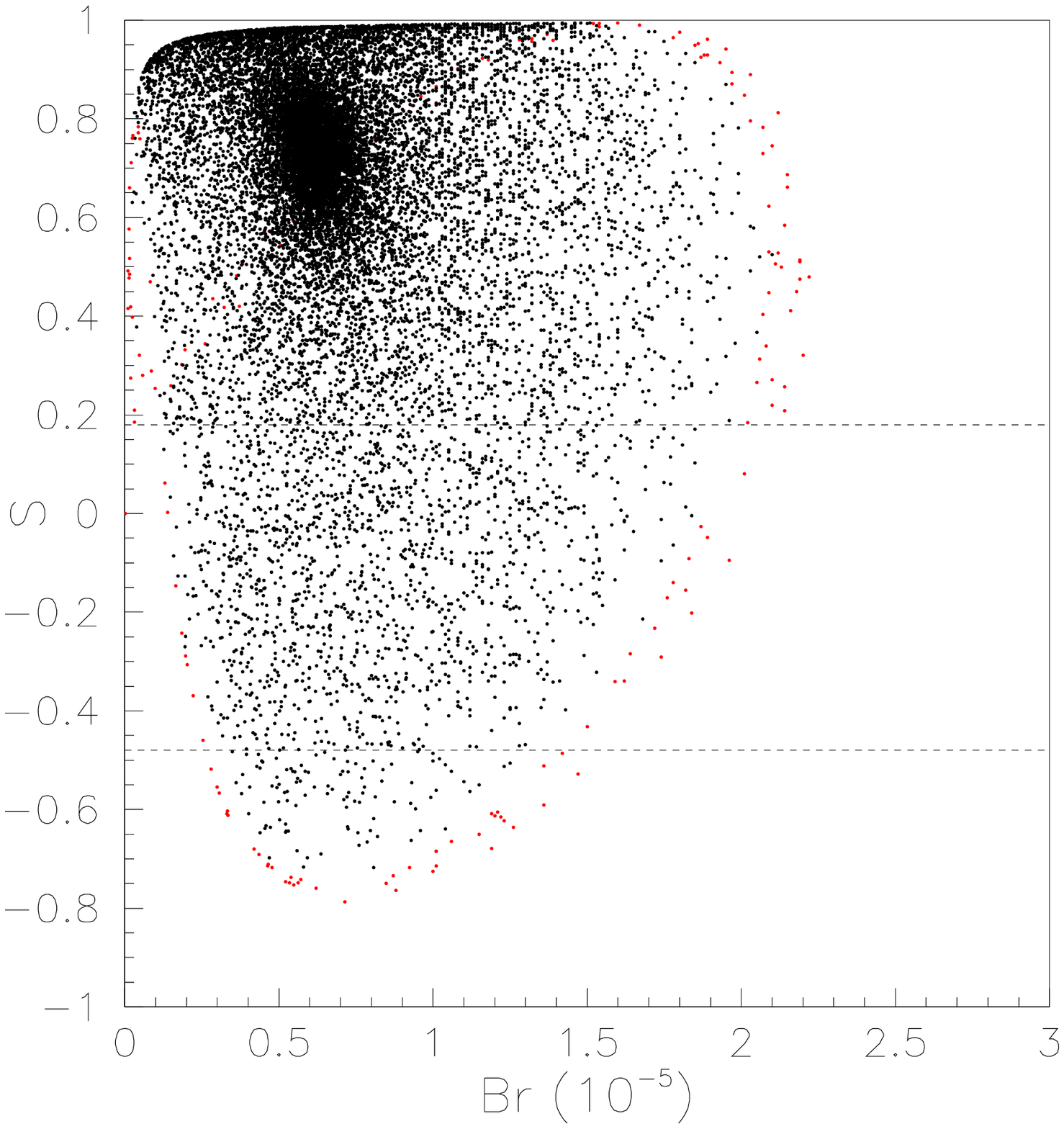}}
{\includegraphics[width=6cm] {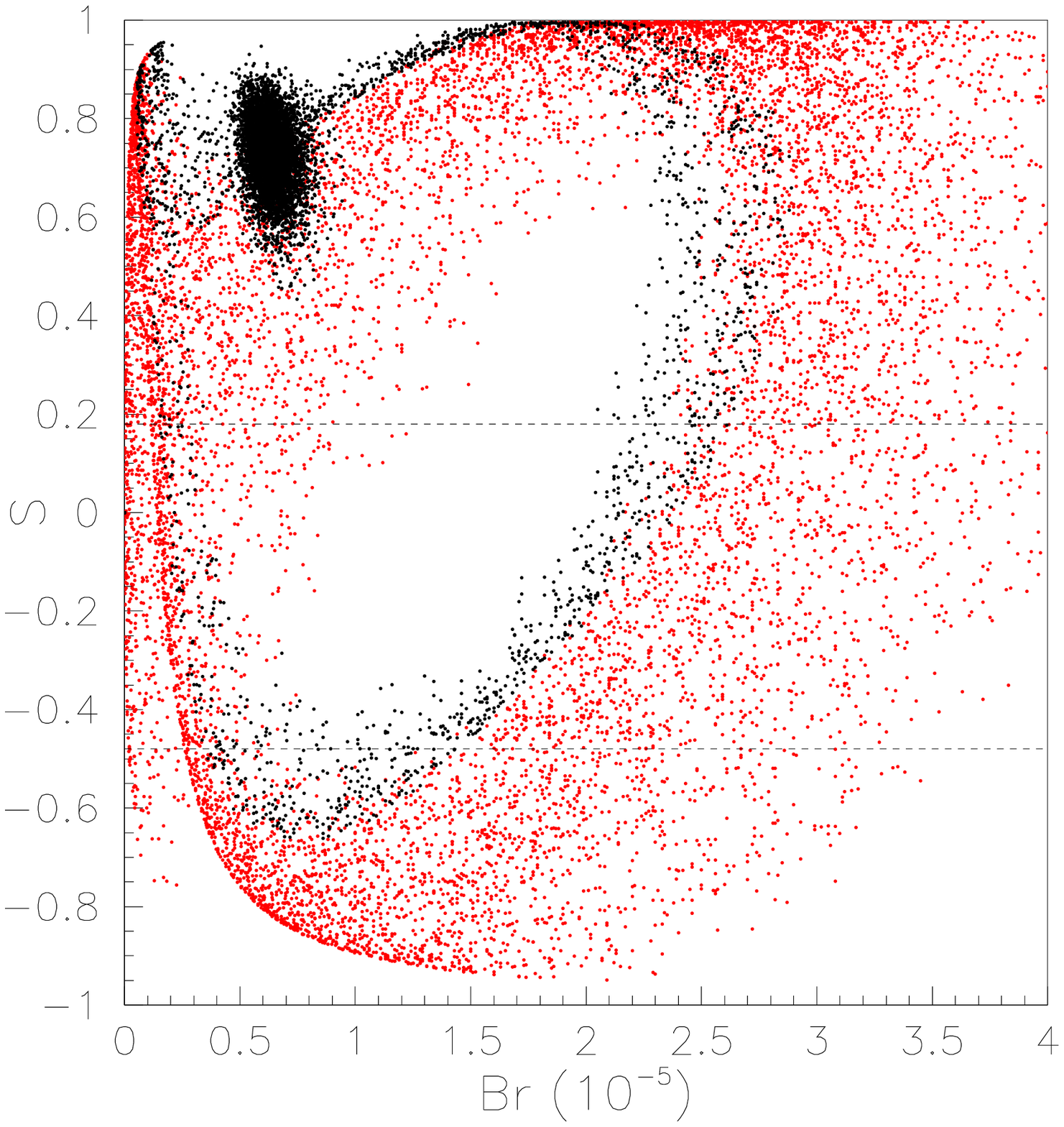}} \\
(a) \hspace{5cm} (b) \caption{
\label{figa0m1000} The correlation between $S_{\phi K_S}$ and
Br($B\to \phi K_S$) for $A_0 = -1000$GeV. The black (red) dots are
allowed by all relevant experimental bounds including (without
including) Br($b \rightarrow s g$). (a) is for the SUSY
contributions with only gluino propagated in the loop,
(b) is for the all contributions included.
Current $1\sigma$ bounds are shown by the dashed lines.
}
\end{figure}

We find from Fig.~\ref{figa0m1000} that in both the cases (a) and
(b) there are regions of parameters where $S_{\phi K_S}$ falls in
$1 \sigma$ experimental bounds and Br is smaller than $1.6\times
10^{-5}$. In the case of all contributions included the region is
larger than that in the case with the contributions from only
gluino-squark in the loop included if the constraint from $b\to s
g$ is not imposed.
There exist two regions of the parameter space as given in
Fig.~\ref{figa0m1000} (b) for the all contributions. The dense
black region corresponds to the heavy SUSY spectrum region with
$m_0, M_{1/2}$ as large as $600 \sim 800$GeV, while the scattered
belt region denotes $M_{1/2}$ as small as $100 \sim 200$GeV and
$m_0$ in the region $500 \sim 800$GeV where sleptons are as heavy
as $500 \sim 800$GeV and Br($\tau \to \mu \gamma$) constrains
$\delta^{dRR}_{23}$ to be order of $0.1$. That Br($b \rightarrow s
g$) serves as a strong constraint on the parameter space is shown
in Fig.~\ref{figa0m1000}, in particular, the Fig.~\ref{figa0m1000}
(b) with all the contributions included. In the scattered belt
region of the Fig.~\ref{figa0m1000} (b), with $M_{1/2}$ as small
as $100 \sim 200$GeV, the destructive chargino contribution to
$C_{7\gamma}$ drives $C_{7\gamma}$ smaller than $C^{\rm
SM}_{7\gamma}$, which calls for larger $C^\prime_{7\gamma}$ of
gluino contributions to enhance Br($b \rightarrow s \gamma)
\propto |C_{7\gamma}|^2 + |C^\prime_{7\gamma}|^2$ above the
experimental lower bound. However large $C^\prime_{7\gamma}$
results in large $C^\prime_{8g}$, which raises Br($b \rightarrow s
g$) beyond the experimental upper bound. Therefore, the most of
the region without the $b \rightarrow s g$ constraint is excluded
when the $b \rightarrow s g$ constraint is imposed. In the case of
$A_0=0$, Fig.~\ref{figa0} (b) corresponds to only the heavy SUSY
spectrum region because the light mass spectrum region, which
corresponds to $100 GeV\leq m_0, M_{1/2}\geq 400 GeV$ roughly, is
excluded by Br($B\to X_s \gamma$) and the experimental lower bound
of $m_{h^0}$ and has no results with $S_{\phi K_S}$ near $0$,
while Fig.~\ref{figa0} (a), which corresponds to the region of the
parameter space with $M_{1/2}$ as small as $100 \sim 200$GeV and
$m_0$ in the region $500 \sim 800$GeV, has some results with
$S_{\phi K_S}$ near $0$, because Br($b \rightarrow s \gamma$)
constraints can be satisfied easily in the latter case. Comparing
Fig.~\ref{figa0m1000} with Fig.~\ref{figa0}, the conclusion is
that the SUSY effects on $S_{\phi K_S}$ for a negative large $A_0$
(say, $|A_0| \gsim m_0, M_{1/2}$) are larger than those for
$A_0=0$. The reason is as follows. One needs to have a large $\mu$
in order to have a large RL mixing between down squarks, i.e., a
large $\delta^{dRL}_{33}$, and consequently a large induced
$\delta^{dRL}_{23}$. Due to the RGE running, a negative large
$A_0$ can drive $\mu$ at the low energy large, whereas $A_0=0$ or
a small $A_0$ fails. And at the same time, a large $\mu$ can make
the chargino contributions significant, which make the real part
of $C_{8g}+C_{8g}^\prime$ smaller than that in the small $A_0$
case\footnote{In the paper we do not consider the cancellation
mechanism~\cite{cm} in the analysis of electric dipole moments
(EDMs) of electron and neutron in SUSY models and assume $\mu,
\;A_0$ are real.}.

We vary $\tan\beta$ and find similar results for $\tan\beta=10,
30$. The numerical results are obtained for $100 GeV \leq
m_0,\;M_{1/2}\leq 800 GeV$. For fixed $m_{\tilde{g}}$, the Wilson
coefficient $C_{8g}^{(\prime)}$ is not sensitive to the variation
of the mass of squark in the range about from $100$GeV to
$1500$GeV. Therefore, the numerical results are not sensitive to
$m_0$ for fixed $M_{1/2}$ and would have a sizable change when
$M_{1/2}$ decreases.

\section{Conclusions and Discussions}
In summary we have calculated the mass spectrum and mixing of
sparticles in the SUSY SO(10) GUT. We have calculated the chargino
contributions to Wilson coefficients at LO using the MIA with
double insertions in SUSY models. Using the Wilson coefficients
and hadronic matrix elements previously obtained, we have
calculated the time-dependent CP asymmetries $S_{\phi k}$ and
branching ratios for the decay $B\rightarrow \phi K_S$. It is
shown that in the reasonable region of parameters where the
constraints from $\tau\to\mu\gamma$, $B_s-\bar{B}_s$ mixing ,
$\Gamma(b \to s \gamma)$, $\Gamma(b \to s g)$, $\Gamma(b \to
s\mu^+ \mu^-)$, and $B\to \mu^+\mu^-$ are satisfied, the branching
ratio of the decay for $B\rightarrow \phi K_S$ can be smaller than
$1.6 \times 10^{-5}$, and $S_{\phi K_S}$ can be negative. In some
regions of parameters $S_{\phi K_S}$ can be as low as $-0.6$.

It is necessary to make a theoretical prediction in SM as
precision as we can in order to give a firm ground for finding new
physics. For the purpose, we calculate the twist-3 and weak
annihilation contributions in SM using the method in
Ref.~\cite{ch} by which there is no any phenomenological parameter
introduced. The numerical results show that the annihilation
contributions to Br are negligible, the twist-3 contributions to
Br are also very small, smaller than one percent, and both the
annihilation and twist-3 contributions to the time-dependent CP
asymmetry are negligible. The conclusion remains in SUSY models
and consequently we neglect the annihilation contributions in
numerical calculations.

In conclusion, we have shown that the recent experimental
measurements on the time-dependent CP asymmetry in $B\to \phi
K_S$, which can not be explained in SM, can be explained in the
SUSY SO(10) grand unification theories where there are flavor
non-diagonal right-handed down squark mass matrix elements of
second and third generations whose size satisfies all relevant
constraints from known experiments ($\tau\to\mu\gamma$, $B\to
X_S\gamma, B_s\to \mu^+\mu^-, B\to X_s \mu^+\mu^-, B\to X_s g,
\Delta M_s$, etc.). Therefore, if the present experimental results
remain in the future, it will signal the significant breakdown of
the standard model and that the SUSY SO(10) GUT is a possible
candidate of new physics.

\section*{Acknowledgement}
The work was supported in part
by the National Nature Science Foundation of China.
XHW is supported by KOSEF Sundo Grant R02-2003-000-10085-0.


\section*{\bf Appendix A~~ Loop functions}
In this Appendix, we present the one-loop function of Wilson
coefficients in this work.
\begin{eqnarray}
F_{2,12,4,34,6,b0}(x) &=&
 x \frac{\partial f_{2,12,4,34,6,b0}(x)}{\partial x} \nonumber\\
F^\prime_{2,12,4,34,6,b0}(x) &=& \frac{x^2}{2}
 \frac{\partial^2 f_{2,12,4,34,6,b0}(x)}{\partial x^2}
\end{eqnarray}
with
\begin{eqnarray}
f_{12}(x) = 9 f_1(x) + f_2(x),
\hspace{1cm} f_{34}(x) = 9 f_3(x) + f_4(x)/2 \nonumber
\end{eqnarray}
where $f_{1,2,3,4,6,b0}$ are given in Ref.~\cite{hw}.


\section*{References}
\begin{thebibliography}{99}
\bibitem{skatm}
Y. Fukuda et al. [Super-Kamiokande Collaboration], Phys. Rev.
Lett. {\bf 81}(1998)1562.
\bibitem{sno}
Q.R. Ahmad et al.  [SNO Collaboration], Phys. Rev. Lett. {\bf
89}(2002)011301; Phys. Rev. Lett. {\bf 89}(2002)011302.
\bibitem{kamland}
K. Eguchi et al. [KamLAND Collaboration], Phys. Rev. Lett. {\bf
90}(2003)021802.
\bibitem{chooz} M. Apollonio et al.[CHOOZ Collaboration],
Phys.Lett.B{\bf 466}(1999)415.
\bibitem{wwzf} S. Weinberg, Trans.N.Y.Acad.Sci.{\bf 38}(1977)185;
F. Wilczek and A. Zee, Phys. Lett. B{\bf 70}(1977)418; H.
Fritzsch, Phys. Lett. B{\bf 70}(1977)436.
\bibitem{FN}
C. D. Froggatt and H. B. Nielsen, Nucl. Phys. {\bf B}147 (1979)
277.
\bibitem{asy} K.S. Babu and S.M. Barr, Phys. Lett. {\bf B381} (1996) 202;
         C.H. Albright, K.S. Babu, and S.M. Barr, Phys. Rev. Lett.
         {\bf 81} (1998) 1167; J. Sato and T. Yanagida,
         Phys. Lett. {\bf B430} (1998) 127; N. Irges, S. Lavignac,
         and P. Ramond, Phys. Rev. D {\bf 58} (1998) 035003.
\bibitem{bdv}W. Buchmuller, D. ~Delepine and F. ~Vissani, Phys. Lett.
{\bf B459} 171 (1999); W. Buchmuller, D. Delepine and L. T.
Handoko, Nucl. Phys.\ {\bf B576} 445 (2000); J. Ellis, M. E.
Gomez, G. K. Leontaris, S. Lola and D. V. Nanopoulos, Eur. Phys.
J. {\bf C14} 319 (2000); J. Hisano and K. Tobe, Phys. Lett. {\bf
B510} 197 (2001); J. A. Casas and A. Ibarra, arXiv:hep-ph/0103065;
D. F. Carvalho, J. Ellis, M. E. Gomez and S. Lola,
arXiv:hep-ph/0103256; T. Blazek and S. F. King,
arXiv:hep-ph/0105005; J. Sato, K. Tobe and T. Yanagida, Phys.
Lett. {\bf B498} 189 (2001); J. Sato and K. Tobe, Phys. Rev. {\bf
D63} 116010 (2001); S. Lavignac, I. Masina and C. A. Savoy,
arXiv:hep-ph/0106245; T.\ Moroi, JHEP.\ {\bf 0003} 019 (2000),
  [arXiv:hep-ph/0002208];  N.\ Akama, Y.\ Kiyo, S.\ Komine and T.\ Moroi,
Phys.\ Rev.\ {\bf D64} 095012 (2001)
  [arXiv:hep-ph/0104263];
T.\ Moroi, Phys.\ Lett.\ {\bf B493}, 366 (2000) [arXiv:
hep-ph/0007328].
\bibitem{lfv3} J. Hisano, T. Moroi, K. Tobe and M. Yamaguchi,
Phys. Rev. D{\bf 53}(1996)2442(hep-ph/9510309); J. Hisano and D.
Nomura, Phys. Rev. D{\bf 59}(1999)116005(hep-ph/9810479).
\bibitem{bi} X-J. Bi, Y-B. Dai and X-Y Qi, Phys. Rev.{\bf D63}
096008 (2001); X-J. Bi and Y-B. Dai, Phys. Rev.{\bf D66} 076006
(2002).
\bibitem{cmm}D. Chang, A. Masiero and H. Murayama, Phys. Rev.
{\bf D67} (2003) 075013.
\bibitem{mvv} A. Masiero, S. K. Vempati and O. Vives,
Nucl. Phys. B{\bf 649}(2003)189(hep-ph/0209303).
\bibitem{bsv}B. Bajc, G. Senjanovi$\acute{c}$ and F. Vissani,
hep-ph/0210207; H.S. Goh, R.N. Mohapatra and S.-P. Ng,
hep-ph/0303055.
\bibitem{hs}J. Hisano and Y. Shimizu, hep-ph/0303071.
\bibitem{hll} C.-S. Huang, T. Li, W. Liao,
 Nucl. Phys. {\bf B673}(2003) 331.
\bibitem{sj}B.~Aubert et al, BABAR Collaboration,
Phys. Rev. Lett. {\bf 89} (2002) 201802; K.~Abe et al, Belle
Collaboration, arXiv:hep-ex/0308036.
\bibitem{2002}Aubert  et al. (BABAR Collaboration),
hep-ex/0207070; T.~Augshev, talk given at ICHEP 2002 (Belle
Collaboration), BELLE-CONF-0232; K. Abe  et al., BELLE-CONF-0201
hep-ex/0207098.
\bibitem{2003}The Belle Collaboration, K. Abe et al,
hep-ex/0308035(BELLE-CONF-0344);
the talk given by T. Browder at LP2003, \\
{\em
http://conferences.fnal.gov/lp2003/program/S5/browder\_s05\_ungarbled.pdf.}
\bibitem{dat}M. B. Causse, hep-ph/0207070; G. Hiller, Phys. Rev.
{\bf D66} (2002) 071502;
A. Datta, Phys. Rev. {\bf D66}(2002) 071702;  M. Raidal, Phys.
Rev. Lett. {\bf 89}(2002) 231803; K. Agashe and C.D. Carone,
hep-ph/0304229; B. Dutta, C.S. Kim, S. Oh, Phys. Rev. Lett. {\bf
90} (2003) 011801;
 J.-P. Lee, K.Y. Lee, hep-ph/0209290; Y.-L. Wu, Y.-F. Zhou,
 hep-ph/0403252.
 \bibitem{hz}C.-S. Huang and S.-H. Zhu, Phys. Rev. {\bf D68} (2003)
 114020[arXiv:hep-ph/0307354].
\bibitem{kk}M. Ciuchini, L. Silvestrini,
 Phys. Rev. Lett. {\bf 89}(2002) 231802;
 L.  Silvestrini, hep-ph/0210031(talk contributed at ICHEP02);
 S.~Khalil, E. Kou, Phys. Rev.  {\bf D67} (2003) 055009;
 R.  Harnik, D.T. Larson, H. Murayama, A. Pierce, hep-ph/0212180;
A. Kundu and T. Mitra, hep-ph/0302123; S.~Khalil, E.~Kou, Phys.
Rev. Lett.
{\bf 91} (2003) 241602;
R. Arnowitt, B. Dutta and B. Hu,  Phys. Rev. {\bf D68} (2003)
075008; J. Hisano, Y. Shimizu, hep-ph/0308255; S. Baek, Phys.
Rev. {\bf D67} (2003) 096004;
 D. Chakraverty, E. Gabrielli, K. Huitu, and S.
Khalil, Phys. Rev. {\bf D68} (2003) 095004.
C. Dariescu, M. A. Dariescu, N.G. Deshpande and D. K. Ghosh,
hep-ph/0308305; Y. Wang, hep-ph/0309290;
 V. Barger, C.-W. Chiang, P. Langacker, H.-S. Lee, hep-ph/0310073;
S. Mishima, A. I. Sanda, hep-ph/0311068; N. G. Deshpande, D. K.
Ghosh, hep-ph/0311332; B. Dutta, C. S. Kim, S. Oh, G. Zhu,
hep-ph/0312389; C.H. Chen, C.Q. Geng, hep-ph/0403188.
\bibitem{kane}G.L. Kane et al., hep-ph/0212092,
 Phys. Rev. Lett. {\bf 90} (2003) 141803.
\bibitem{msd}A. Stocchi, Nucl. Phys. Proc. Suppl. {\bf 117}(2003) 145
[arXiv:hep-ph/0211245].
\bibitem{chw}J.-F. Cheng, C.-S. Huang and X.-H. Wu, Phys. Lett. {\bf B 585}
 (2004) 287 [arXiv:hep-ph/0306086].
\bibitem{hlm}R. Harnik, D.T. Larson, H. Murayama, A. Pierce,
hep-ph/0212180.
\bibitem{cmsvv}M. Ciuchini et al., Phys. Rev. Lett. {\bf 92} (2004)
071801.

\bibitem{taumugamma}K.Abe et. al., Belle collaboration,
Phys. Rev. Lett. {\bf 92} (2004) 171802 [hep-ex/0310029].

\bibitem{hlmp}R. Harnik et al., hep-ph/0212180.
\bibitem{hw}C.-S. Huang and X.-H. Wu, Nucl. Phys. {\bf B657}(2003) 304.
\bibitem{mi}L.~J.~Hall, V.~A.~Kostelecky, S.~Raby, Nucl. Phys. {\bf B267}
(1986) 415.
\bibitem{gabbiani}For a review, see: F.~Gabbiani, E.~Gabrielli, A.~Masiero,
L.~Silvestrini, Nucl. Phys. {\bf B477} (1996) 321.
\bibitem{li1}H.-n.~Li and H.~Yu, Phys. Rev. Lett. {\bf 74} (1995) 4388;
H.-n.~Li and T.~Yeh, Phys. Rev. {\bf D56} (1997) 1615.
\bibitem{li}Y.~Y.~Keum, H.-n.~Li and A.~I.~Sanda,
Phys. Lett. {\bf B504} (2001) 6; Phys. Rev. {\bf D63} (2001)
054008.
\bibitem{bbns}M. Beneke et al., Phys. Rev. Lett. {\bf 83}(1999) 1914;
 Nucl. Phys. {\bf B591}(2000) 313.
\bibitem{bbns1}M. Beneke et al., Nucl. Phys. {\bf B606}(2001) 245.

\bibitem{nir}Y. Nir, Nucl. Phys. Proc. Suppl. {\bf 117} (2003)
111.
\bibitem{dh}S. Dimopoulos and L.J. Hall, Phys. Lett. {\bf B344},
185 (1995).
\bibitem{seesaw} M. Gell-Mann, P. Rammond and R. Slansky, in {\it
Supergravity}, eds. D. Freedman {\it et al.} (North-Holland,
Amsterdam, 1980); T. Yanagida, in proc. KEK workshop, 1979
(unpublished); R.N. Mohapatra and G. Senjanovi\'c, Phys. Rev.
Lett. {\bf 44}, 912 (1980); S. L. Glashow, {\it Cargese lectures},
(1979).
\bibitem{bw}W. Buchmueller and D. Wyler, Phys. Lett. {\bf B521},
291 (2001).
\bibitem{bur}G.~Buchalla, A.~J.~Buras and M.~E.~Lautenbacher,
Rev.\ Mod.\ Phys.\  {\bf 68}, 1125 (1996) [arXiv:hep-ph/9512380].
\bibitem{everett01}L.~Everett, G.~L.~Kane, S.~Rigolin, L.~-T.~Wang, and
T.~T.~Wang, JHEP 0201 (2002) 022.

\bibitem{lunghi2000}E.~Lunghi, A.~Masiero, I.~Scimemi, and L.~Silvestrini,
Nucl. Phys. {\bf B568} (2000) 120.

\bibitem{adm}J.A. Bagger, K.T. Matchev and R.J. Zhang, Phys. Lett.
{\bf B412}(1997) 77; M. Ciuchini et al., Nucl. Phys. {\bf
B523}(1998) 501; C.-S. Huang and Q.-S. Yan, "The frontiers of
physics at millennium" (Beijing 1999), Editor Y.-L. Wu, p.129
[hep-ph/9906493]; A.J. Buras, M. Misiak and J. Urban, Nucl.Phys.
{\bf B586} (2000) 397.

\bibitem{bghw}F. Borzumati, C. Greub, T. Hurth and D. Wyler, Phys.
Rev. {\bf D62}(2000) 075005 .

\bibitem{hk}G. Hiller, F. Kr$\ddot{u}$ger, hep-ph/0310219.
\bibitem{chw1}J.-F. Cheng, C.-S. Huang and X.-H. Wu,
hep-ph/0404055.
\bibitem{hmw}X.-G. He, J.P. Ma and C.-Y. Wu, Phys. Rev.
{\bf D63} (2001) 094004; H.-Y. Cheng and K.C. Yang, Phys. Rev.
{\bf D64}(2001) 074004.
\bibitem{Keum:2000ph}
Y.~Y.~Keum, H.~n.~Li and A.~I.~Sanda,
Phys.\ Lett.\ B {\bf 504}, 6 (2001) [arXiv:hep-ph/0004004].
\bibitem{gt}B.V. Geshkenbein and M.V. Terentev, Phys. Lett. {\bf
117}, 243 (1982).
\bibitem{fks}
T.~Feldmann, P.~Kroll and B.~Stech,
Phys.\ Rev.\ D {\bf 58}, 114006 (1998)
[hep-ph/9802409];\\
T.~Feldmann,
Int.\ J.\ Mod.\ Phys.\ A {\bf 15}, 159 (2000)
[hep-ph/9907491].
\bibitem{Kaiser:1998ds}
R.~Kaiser and H.~Leutwyler,
Proceedings of Workshop on Nonperturbative Methods in
Quantum Field Theory, Adelaide, 1998 [hep-ph/9806336];\\
R.~Kaiser and H.~Leutwyler,
Eur.\ Phys.\ J.\ C {\bf 17} (2000) 623
[hep-ph/0007101].
\bibitem{ciuchini}M. Ciuchini {\it et al}, J. High Energy Phys. {\bf
07}(2001), 13

\bibitem{chenghy}Hai-Yang Cheng, Yong-Yeon Keum and Kwei-Chou
Yang, Phys.Rev.{\bf D65} (2002) 094023.

\bibitem{bsmu}D. Acosta et al. (CDF Collaboration),
hep-ex/0403032.
\bibitem{pdg}K. Hagiwara et al., Phys. Rev. {\rm D66}, 010001 (2002).
\bibitem{nanop}J.L. Lopez, D.V. Nanopoulos, X. Wang and A.
Zichichi, Phys. Rev. {\bf D51}, 147 (1995).
\bibitem{hy}Y.-B. Dai, C.-S. Huang and H.-W. Huang,
Phys. Lett. {\bf B390}(1997) 257;
C.-S. Huang and Q.-S. Yan, Phys. Lett. {\bf B442}(1998) 209;
C.-S. Huang, W. Liao, and Q.-S. Yan, Phys. Rev. {\bf D59}(1999) 011701.
\bibitem{chlw} N. Chamoun, C.-S. Huang, C. Liu, X.-H. Wu, Nucl. Phys.
{\bf B624} (2002) 81.
\bibitem{bsg}A. Kagan, hep-ph/9806266; T.E. Coan et al. (CLEO
Collaboration), Phys. Rev. Lett. {\bf 80}, 1150 (1998).
\bibitem{murayama0212}R.~Harnik, D.~T.~Larson, H.~Murayama, A.~Pierce,
arXiv:hep-ph/0212180.
\bibitem{ch}J.-F. Cheng and C.-S. Huang, Phys. Lett. {\bf B554}(2003) 155
. In the paper the constraint from $B\to \mu^+\mu^-$ is not
imposed. \bibitem{datanew}D. Acosta et al. (CDF Collaboration),
hep-ex/0403032. \bibitem{cm}T.Ibrahim and P.Nath, Phys.Rev. D{\bf
57},478(1998); (E) ibid, {\bf D58}, 019901(1998); Phys. Rev. {\bf
D58},111301(1998); M.Brhlik,G.J.Good,and G.L.Kane, Phys. Rev. {\bf
D59},11504(1999); C.-S. Huang and Liao Wei, Phys. Rev. {\bf D61}
(2000) 116002; ibid. {\bf D62} (2000) 016008.
\end{thebibliography}
\end{document}